\RequirePackage{fix-cm}
\documentclass[smallcondensed]{svjour3}     
\smartqed  
\usepackage{amsmath}
\usepackage{graphicx}
\usepackage{tabularx}
\usepackage{longtable}
\usepackage{caption}
\usepackage{tabularray}
\usepackage{multirow}
\usepackage{booktabs}
\usepackage[hyphens,spaces,obeyspaces]{url}
\PassOptionsToPackage{hyphens}{url}
\usepackage[hidelinks]{hyperref}
\hypersetup{breaklinks=true}
\usepackage[round]{natbib} 
\usepackage[dvipsnames]{xcolor}
\usepackage{tcolorbox}
\usepackage[colorinlistoftodos, textsize=small]{todonotes}
\hypersetup{
    colorlinks,
    linkcolor={red!50!black},
    citecolor={blue!50!black},
    urlcolor={blue!80!black}
}
\usepackage{marginnote}

\definecolor{pitchblack}{cmyk}{1 1 1 1}

\newcommand{\securityIssue}[1]{security issue#1}

\newcommand{\vulnerability}[1]{vulnerabilit#1}

\newcommand{\knownV}{known}
\newcommand{\knownVC}{Known}
\newcommand{\weaknessC}[1]{Coding weakness#1}
\newcommand{\securityConcernC}[1]{Security concern#1}
\newcommand{\weakness}[1]{coding weakness#1}
\newcommand{\securityConcern}[1]{security concern#1}

\newcommand{\rqOne}{What kinds of security concerns related to coding weaknesses are often raised in code review?}
\newcommand{\rqTwo}{How aligned are the raised security concerns and \knownV{} vulnerabilities?}
\newcommand{\rqThree}{How are security concerns handled in code review?}
\newcommand{\smallheading}[1]{\vspace{2mm}\noindent\textbf{#1} \\}
\newcommand{\inlineheading}[1]{\vspace{2mm}\noindent\textbf{#1}}
\newcommand{\leftalignedcaption}[1]{%
  \captionsetup{justification=raggedright,singlelinecheck=false,labelfont=bf}%
  \caption{#1}%
}
\newcommand{\reviewersComment}[1]{\marginnote{\textcolor{ForestGreen}{}}} 
\newcommand{\reviewersModification}[1]{\textcolor{pitchblack}{#1}} 

\begin{document}

\title{
%
Toward Effective Secure Code Reviews: An Empirical Study of Security-Related Coding Weaknesses

}

\titlerunning{Toward Effective Secure Code Reviews}

\author{Wachiraphan Charoenwet         \and
        Patanamon Thongtanunam \and
        Van-Thuan Pham \and
        Christoph Treude
}


\institute{Wachiraphan Charoenwet \at
              wcharoenwet@student.unimelb.edu.au 
           \and
           Patanamon Thongtanunam \at
              patanamon.t@unimelb.edu.au
           \and
           Van-Thuan Pham  \at
              thuan.pham@unimelb.edu.au
            \and
            School of Computing and Information Systems \\
            Faculty of Engineering and Information Technology\at
            The University of Melbourne, Australia
           \and
           Christoph Treude \at
              ctreude@smu.edu.sg
            \and
            School of Computing and Information Systems\at
            Singapore Management University, Singapore
}


\maketitle

\begin{abstract}
Identifying \securityIssue{s} early is encouraged to reduce the latent negative impacts on the software systems.
Code review is a widely-used method that allows developers to manually inspect modified code, catching \securityIssue{s} during a software development cycle. 
However, existing code review studies often focus on \knownV{} \vulnerability{ies}, neglecting \weakness{es}, which can introduce real-world \securityIssue{s} that are more visible through code review. 
The practices of code reviews in identifying such \weakness{es} are not yet fully investigated.

To better understand this, we conducted an empirical case study in two large open-source projects, OpenSSL and PHP.
Based on 135,560 code review comments, we found that reviewers raised \securityConcern{s} in 35 out of 40 \weakness{} categories. Surprisingly, some \weakness{es} related to past \vulnerability{ies}, such as memory errors and resource management, were discussed less often than the \vulnerability{ies}. 
Developers attempted to address raised \securityConcern{s} in many cases (39\%-41\%), but a substantial portion was merely acknowledged (30\%-36\%), and some went unfixed due to disagreements about solutions (18\%-20\%).
This highlights that \weakness{es} can slip through code review even when identified. 
Our findings suggest that reviewers can identify various \weakness{es} leading to \securityIssue{s} during code reviews.
However, these results also reveal shortcomings in current code review practices, indicating the need for more effective mechanisms or support for increasing awareness of \securityIssue{} management in code reviews.

\end{abstract}

\section{Introduction}
\label{intro}
Software security is an important focus in software development processes because it encompasses how software system sustains external threats~\citep{McGraw2004SoftwareSecurity}. Managing \securityIssue{s} in software products is crucial because the latent \securityIssue{s}, especially exploitable \vulnerability{y}, can exponentially impact end-users and require more resources to resolve if discovered in the later stage.
Attempting to mitigate \securityIssue{s}, developers are encouraged by the ongoing \textit{shift-left} concept~\citep{Migues2021WhyTesting, Weir2022ExploringResponsibility} to test the new software as early as possible. 
In the spirit of shifting left, numerous organizations have adopted modern code review, a software quality assurance activity for identifying and removing software defects early in the development lifecycle~\citep{Bosu2013ModelingOrganizations, Rigby2012ContemporaryDevelopment}. Prior studies reported that code review is a potential approach for identifying and eliminating \securityIssue{s}~at the early stage~\citep{Hein2009SecureChallenges, Bosu2014IdentifyingStudy, Assal2018SecurityLifecycle}. In particular, a study by~\cite{DiBiase2016AChromium} observed that code review could identify well-known \securityIssue{s} 
such as Cross-Site Scripting (XSS).


Several studies have investigated the benefits of code reviews in identifying \securityIssue{s}~\citep{Alfadel2023EmpiricalPackages, Bosu2014IdentifyingStudy, DiBiase2016AChromium, Edmundson2013AnReview, Paul2021WhyProject}. Still, the \securityIssue{s} studied by previous works were typically bounded by the types of well-known vulnerabilities such as SQL Injection and XSS. In particular, the majority of studied \securityIssue{s} are limited to the \vulnerability{ies} that attackers can exploit.
Since code review focuses on identifying and mitigating coding issues, we hypothesize that \weakness{es}, or faults in code, that can potentially lead to \securityIssue{s} may also be found and mitigated during the code review process.
\reviewersComment{R2.3, R2.5}
\reviewersModification{
Moreover, \weakness{es} should fit the capability of the typical reviewers who may have limited security awareness and knowledge~\citep{Braz2022LessFormat} because reviewers need to have a substantial understanding of security knowledge in order to identify the \securityIssue{s} during code review~\citep{Braz2022SoftwarePerspective}.
Our preliminary analyses (Section~\ref{preliminary_analysis}) indicate that the simple \weakness{es} such as numeric errors, insufficient input validation, or business logic errors are more frequently discussed by reviewers than the security issues that prior work regularly studied in code reviews.
}
However, the practices of code reviews in identifying such \weakness{es} are not yet fully investigated. 
This includes the types of \weakness{es}~that lead to \securityIssue{s}~and the handling process of these \weakness{es}.
In addition, little is known about whether the security concerns raised during code reviews are aligned with the \vulnerability{ies} that a system may have had in the past.

Exploring these aspects would help us better understand the unrealized benefits of considering \weakness{es} during code reviews for the early prevention of software \securityIssue{s}. Such insight can also reveal the gaps between the current code review practice and the \vulnerability{ies} that were \knownV{} in the respective systems. On one hand, software teams could develop secure code review policies that enable them to more effectively identify and address security concerns during the code review process~\citep{Mantyla2009WhatReviews}. On the other hand, researchers can expand the new perspective of code review studies and understand the shortcomings in code review practices and tools.

In this work, we aim to investigate the \weakness{es} that were raised during the code review and to investigate how the code review comments that mentioned \weakness{es} were handled.
We conducted our case study on OpenSSL and PHP which are large open-source systems that are prone to \securityIssue{s}.
\reviewersComment{R2.3}
\reviewersModification{
We chose to examine these phenomena in open-source projects due to the availability of publicly accessible datasets, the mandatory code review policy, and the past vulnerabilities of the selected projects. 
This decision also stems from the observation that code review outcomes in open-source communities, such as the ratio between functional defects and maintainability defects identified by reviewers, do not significantly differ from those observed in industry settings.~\citep{Mantyla2009WhatReviews, Beller2014ModernFix}.}
To confirm our presumption that the discussion related to \weakness{es} is more prevalent than the explicit \vulnerability{ies}, we conducted an initial analysis by manually annotating 400 randomly sampled code review comments from each studied project.
We found that \weakness{es} could be raised in the code reviews 21 - 33.5 times more often than explicit \vulnerability{ies}. 

Therefore, we conducted an empirical study to address three research questions: \textbf{(RQ1)} \textit{\rqOne{}}, \textbf{(RQ2)} \textit{\rqTwo{}}, and \textbf{(RQ3)} \textit{\rqThree{}}.
To do so, we applied a semi-automated approach to 135,560 code review comments to identify code review comments that are related to \weakness{es}.
Then, we manually annotated the types of \weakness{es} for 6,146 code review comments that are related to \weakness{es}.
We used the taxonomy of the Common Weakness Enumeration---CWE-699 which covers 40 categories of \weakness{es} that are related to \securityIssue{s}. 
In addition, we analyzed 378 Common Vulnerabilities Exposure (CVE) reports (101 from OpenSSL; 277 from PHP) to investigate whether the \weakness{es} raised during the code reviews aligned with the \knownV{} \vulnerability{ies} in the systems. 
To understand how \weakness{} comments were handled during code reviews, we performed qualitative analysis to identify the handling scenarios based on the code review activities (e.g., review discussion, revisions)  of the corresponding code changes. 

The case study results show that \weakness{es} related to 35 out of the 40 categories in CWE-699 were raised during the code review process of OpenSSL and PHP (RQ1).
For example, comments about \weakness{es} in authentication, privilege, and API were frequently raised in both studied projects. Each studied project also has unique \weakness{es} raised during code reviews, e.g., the direct security threats in OpenSSL and input data validation in PHP. 
These results indicate that various \weakness{es} that link to \securityIssue{s} were raised during the code review process, and the different software projects have different focused \weakness{es}.
\knownVC{} \vulnerability{ies} in the studied projects are related to 16 weakness categories (RQ2). However, \weakness{es} related to memory buffer errors and resource management errors are the least frequently discussed coding weaknesses in OpenSSL and PHP (4\%-9\%), despite the high percentages of known vulnerabilities (17\%-29\%). 
These results suggest that some important \weakness{es} in a project may not be sufficiently discussed in the current code review practice.

\weaknessC{es} raised during the code reviews were handled in four ways (RQ3). In many cases (39\%-41\%), developers attempted to solve the issues. Nevertheless, approximately a third (30\%-36\%) of the raised \weakness{es} were only acknowledged without immediate fixes (i.e., no additional modifications to the code changes in the reviews). We observed that some of the acknowledged concerns were agreed to be fixed in new separate code changes (10\%-18\%) and some were left without fixing due to disagreement about the proper solution (18\%-20\%). A relatively small proportion of the concerns raised (14\%-26\%) were clarified and dismissed through discussion. From all scenarios, we found alarming cases (6\%-9\%) where \securityIssue{s} can be introduced in the code because code changes with unresolved discussion were eventually merged. Additionally, the abandoned concerns (3\%-9\%) and the unsuccessfully fixed concerns (2\%-4\%) are also important because they can negatively affect the developer's contribution~\citep{Gerosa2021TheSource}. These scenarios indicate that \securityConcern{s} from \weakness{es} need a better handling process. \\


Based on the findings, we recommend the software projects to consider \weakness{} categories (i.e., CWE-699) as a guideline for identifying \weakness{es} that can introduce \securityIssue{s} in code reviews. \weaknessC{es} can be prioritized based on the significance, proneness, and unique set of \weakness{es} raised the past code reviews. 
Our work also highlights a shortcoming of the code review process in handling \securityConcern{s} (i.e., unsuccessfully fixed, unresolved discussion, and unresponded) which might require future work to address. \\


\noindent \textbf{Novelty \& Contributions}:
\reviewersComment{R2.1}
\reviewersModification{To the best of our knowledge, this paper is the first to empirically investigate code review in identifying and mitigating \weakness{es} that link to \securityIssue{s} in addition to the well-known \vulnerability{ies}, highlighting the potential benefits of code reviews for early prevention of potential \securityIssue{s}. 
Second, we presented a novel semi-automated approach that leverages a domain-specific pre-trained word embedding model to find potential code review comments related to \securityIssue{s}.
Third, we examined the alignment between the \knownV{} \vulnerability{ies} in the studied systems and the \weakness{es} that were often raised during the code review process, highlighting a shortcoming of the current code review practices that some important \weakness{es} may not be sufficiently discussed.
Finally, we investigated the handling process of the \weakness{es} raised during a code review which sheds light on an issue that a \weakness{} can slip through the code review process and potentially become a \securityIssue{} in the future.\\
}

\noindent \textbf{Data Availability}: We have released the supplementary material\footnote{\url{https://figshare.com/s/ca1effdc5a9db4f50133}} of scripts for data retrieval and data analysis in this study along with the annotated data to facilitate further research.\\

\noindent \textbf{Paper Organization}:
Section \ref{background} describes the background and explores the related work. 
Section \ref{motivation} demonstrates the examples of vulnerabilities that were caused by \weakness{es}.
Section \ref{design} explains the case study design. Section \ref{preliminary_analysis} explains the initial analysis method and result. Section \ref{results} reports the results. Section \ref{discussion} discusses the implications and suggestions. Section \ref{threats} clarifies the threats that may affect the validity of this study. Finally, Section \ref{conclusion} draws the conclusion.

\section{Background and Related Work} 
\label{background}
In this section, we provide background on software security and the modern code review process. We then discuss the current practice and remaining challenges in code review for identifying \securityIssue{s}.

\subsection{Software Security} 


Software security represents the concepts that help software products survive and operate normally under harmful attacks~\citep{McGraw2004SoftwareSecurity}. 
Software \securityIssue{s} represent a variety of problems related to software security.
Software vulnerability, a type of \securityIssue{}, is a "\textit{.. security flaw, glitch, or weakness found in software code that could be exploited by an attacker ..}"~\citep{Dempsey2020NISTIRManagement}. 
In particular, \cite{Munaiah2017DoProject} explained that the flaws or faults in software can result in \textit{"the violation of the system’s security policies"}. The same study also suggested that these kinds of faults should be prioritized and eliminated as early as possible, as they could constitute considerable impacts in the later stages. 
 

\reviewersModification{
\subsection{Coding Weaknesses}
\label{coding_weakness_definition}
\reversemarginpar\reviewersComment{R2.3, R3.1}
Software \securityIssue{s} can be caused by the chain of common software development faults~\citep{Hoole2016ImprovingAssurance}, so-called \weakness{es}, such as improper data validation, miscalculation, or incorrect memory allocation.
\cite{Bojanova2016TheBugs} introduced the \textit{Bugs Framework} that defines the relationship between the causes (\weakness) and the consequences of various software \securityIssue{s}. Each \securityIssue{} can originate from several causes, influenced by different attributes, and result in various consequences that could be exploited by an attacker. \cite{Bojanova2023BugVulnerabilities} elaborated the \textit{Bugs Framework} by showing several real-world examples of how a sequence of \textit{\weakness{es}} can cause a \securityIssue{}.
To give an example, CVE-2021-21834\footnote{\url{https://nvd.nist.gov/vuln/detail/CVE-2021-21834}} presents a vulnerability in a MPEG-4 library. An attacker can exploit this vulnerability by developing a special input that triggers the buffer overflow and causes a denial of service error.
Although the buffer overflow is part of the vulnerable consequence, the actual cause of the vulnerability is a \weakness{} i.e., improper input validation.
Similarly, \cite{Tsipenyuk2005SevenErrors} also identified seven groups of faults in source code that can compromise software security. For example, \textit{Path Manipulation error} and \textit{Illegal Pointer Value error} belong to the group \textit{Input Validation and Representation} which can lead to \securityIssue{s} such as cross-site scripting (XSS) or SQL Injection.
}

\subsection{Security Shift-Left}
The \textit{Shift-Left} concept was introduced by  \cite{Smith2001Shift-LeftTesting} to motivate practitioners to test the developed software products as early, and from as many aspects, as possible in order to diminish the potential defects that may introduce unexpected consequences in later stages.
\cite{Carter2017FrancoisDevSecOps} suggested \textit{security shift-left} where security awareness should be incorporated into the software development lifecycle at the early stage, allowing the incubation of synergies between security experts and software developers. The early detection of \securityIssue{s} also means smaller cost and effort required to fix them~\citep{McConnell2004CodeEdition}, as well as reducing the impacts on the end-users. 
  
In practice, practitioners can use various approaches to perform \textit{security shift-left}. From the organizational perspective, \cite{Weir2022ExploringResponsibility} provided initial guidelines for companies that want to shift security to the left. 
At the management level, the companies can establish security governance, create a centralized \textit{software security group} which consists of security specialists, and allocate a security satellite team to support security activities across the organization. At the project level, the project team should lead the technical aspect of security matters and request security support from the satellite team when needs arise. 

\subsection{Modern Code Review}
Lightweight code review has been increasingly adopted in the software industry as a replacement for traditional code inspection activity \citep{Rigby2012ContemporaryDevelopment}. A developer who composes the code changes (so-called, Author) submits the changes to the code review system. Then other developers in the project (so-called, Reviewers) can freely review the changes and provide feedback. Although code review is used for multiple purposes such as knowledge transfer or sharing of code ownership, the main objective of the activity remains to manage software defects \citep{Bacchelli2013ExpectationsReview, Rigby2013ConvergentPractices, Thongtanunam2015InvestigatingSystem}. 

\cite{Balachandran2013ReducingRecommendation} presented various factors that affect code reviews such as the experience and available effort of reviewers. Similarly, \cite{Bacchelli2013ExpectationsReview} reported that different levels of source code understanding among code developers affect the quality of code review feedback. In addition, the outcomes of the code review were also discussed. \cite{Mantyla2009WhatReviews} identified two defect classes (1) \textit{functionality} and 2) \textit{evolvability}) that can be discovered during code reviews of nine industrial software projects. The study of \cite{Beller2014ModernFix} confirmed a similar notion based on the ConQAT and GROMACS open-source projects. In addition to these two main types of defects, \cite{Thongtanunam2015InvestigatingSystem} found that code developers may also raise concerns related to traceability (i.e., the bookkeeping of code changes). 
Nevertheless, few studies have delved into the non-functional, yet important defects like \securityIssue{s}.

\subsection{Code Review for Software Security}
\reviewersComment{R2.2}
\reviewersModification{
During the code review process, reviewers may follow the security code review practice~\citep{Howard2006AReviews} to detect various potential \securityIssue{s} in the code.
}
Prior studies have investigated the well-known \securityIssue{s} such as overflow, cross-site scripting, SQL injection, and cross-site request forgery that can be identified during the code review process~\citep{Paul2021WhyProject, Bosu2014IdentifyingStudy, Bosu2013PeerEvaluation, Edmundson2013AnReview, Alfadel2023EmpiricalPackages, DiBiase2016AChromium}. 
\cite{Edmundson2013AnReview} conducted a controlled study by asking developers to identify the injected vulnerable code in a web application called Anchor CMS. 
\cite{Paul2021WhyProject} and \cite{DiBiase2016AChromium} examined code review comments in Chromium projects.
A recent study by \cite{Alfadel2023EmpiricalPackages} examined code review comments in NodeJS packages. 
In brief, they reported similar results that developers can identify certain security issues such as cross-site scripting (XSS), but the effectiveness of \securityIssue{} identification in code reviews is relatively low. 

It can be seen that the types of security issues studied in prior works~\citep{Paul2021WhyProject, Bosu2014IdentifyingStudy, Bosu2013PeerEvaluation, Edmundson2013AnReview, Alfadel2023EmpiricalPackages, DiBiase2016AChromium}  are still limited to the aspects of consequences from the external threats, i.e., how a threat agent (or an attacker) can exploit the system (e.g., cross-site scripting and SQL injection). To be able to identify such external threats during code reviews, reviewers are required to possess a deep understanding of security knowledge~\citep{Braz2022SoftwarePerspective} and endure high cognitive load~\citep{Goncalves2022DoLoad}. 

\reversemarginpar\reviewersComment{R2.2}
\reviewersModification{
Rather than focusing on well-known \securityIssue{s} that are limited to the aspects of consequences from external threats or the visible \vulnerability{ies}, little work has investigated \weakness{es} that potentially introduce \securityIssue{s} and \vulnerability{ies}. 
Since code reviews focus on identifying coding issues in source code, identifying \weakness{es} that can lead to various exploitable \vulnerability{ies} would be more aligned with code review practices. 
Therefore, investigating the types of \weakness{es} will help us better understand the benefits and capabilities of code reviews for software security, as well as highlight any deficiencies in current discussions on \weakness{es} within code reviews. 
Table \ref{table:comparing-related-works} shows the key differences (i.e., studied taxonomy and methodology) between this work and the related secure code review studies.
}


\begin{table}[h]
\centering
\small
\leftalignedcaption{Key differences between prior secure code review studies and this work.}
\begin{tabularx}{\linewidth}{
            >{\hsize=0.9\hsize}X
            >{\hsize=1\hsize}X
            >{\hsize=1\hsize}X
            >{\hsize=0.8\hsize}X
            >{\hsize=1.3\hsize}X
          }
\toprule
\textbf{Study} & \textbf{Studied Systems}  & \textbf{Programming Languages}  & \textbf{Studied Taxonomy} &\textbf{Identification Approach}\\
\hline
 \midrule
    \cite{Alfadel2023EmpiricalPackages} & 10 JavaScript libraries e.g., Electron, React, and Sequelize & JavaScript, TypeScript & 14 Security issues & Keyword-based approach \\
    \hline
    \cite{Paul2021WhyProject} & Chromium OS & C++ &  15 CWE items & Keyword-based approach \\
    \hline
    \cite{DiBiase2016AChromium} & Chromium Browser & C++ &  24 Security issues & Random sampling \\
    \hline
    \cite{Bosu2014IdentifyingStudy}	& 10 open source projects e.g., MediaWiki, Chromium OS, and Qt & C, C++, Java, PHP, JavaScript & 10 CWE items & Keyword-based approach \\
    \hline
    \cite{Edmundson2013AnReview} & 2 Web applications i.e., Anchor CMS
 and oDesk & PHP &  3 Security issues & Developer interview \\
 \hline
    This work	& OpenSSL and PHP & C, C++ & 	40 CWE categories$~{\dag}$ & Semi-automated (similarity ranking) \\
 \bottomrule
 \multicolumn{3}{l}{\tiny ${\dag}$ comprising more than 400 CWE items.}
\end{tabularx}
\newline
\label{table:comparing-related-works}
\end{table}

\subsection{Security Concern Handling Process in Code Review}
To better understand the effectiveness of code reviews, prior studies have investigated how raised concerns were addressed during code reviews. 
\cite{Assal2018SecurityLifecycle} interviewed developers to understand their perception of security throughout the software development lifecycle. While the results show that security can be seen as a \textit{formal checkpoint} in code reviews, the results did not elaborate on how the developers would respond to the \securityConcern{s} raised by reviewers. A study by~\cite{Kononenko2016CodeIt} reported that \securityConcern{s} are one of the aspects that developers consider before making changes. 
In particular, \cite{Han2021UnderstandingCommunity} investigated code-smell-related comments (e.g., violation of coding conventions) in code reviews and reported that 6\% of comments were ignored by developers.
Similarly, \cite{Beller2014ModernFix} 
found that 7\%-35\% of the code review comments were neglected by the developers in general.

Although several studies have investigated the aspects of the code review process, little work has studied how developers respond to concerns about \weakness{}, given that they are related to potential \securityIssue{s}, during code reviews. 
\reviewersComment{R3.1}
\reviewersModification{Referring to Section~\ref{coding_weakness_definition}, it becomes apparent that \weakness{es} can overshadow both functionality and evolvability defects that reviewers can identify during code reviews~\citep{Mantyla2009WhatReviews}.}
An extended understanding of how developers respond to such \weakness{es} could shed light on current secure code review practices and their challenges. Such an understanding would help practitioners develop secure code review policies that address current challenges, allowing the team to execute better secure code reviews and prevent \securityIssue{s} in software products.

\section{Motivating Examples}
\label{motivation}
In this section, we provide motivating examples of \securityIssue{s} that are related to \weakness{es} which can potentially be identified by code reviews. 
We obtained three examples from the Common Vulnerabilities Exposure (CVE) reports for our studied systems (i.e., OpenSSL and PHP). 
A CVE report provides a description of the \knownV{} \vulnerability{ies} and related information including corresponding code changes (or patches), severity score, or related weaknesses that are assigned by the National Vulnerability Database~\citep{NVDProcess} security analysts. \\



\begin{figure}[h]
  \includegraphics[width=\textwidth]{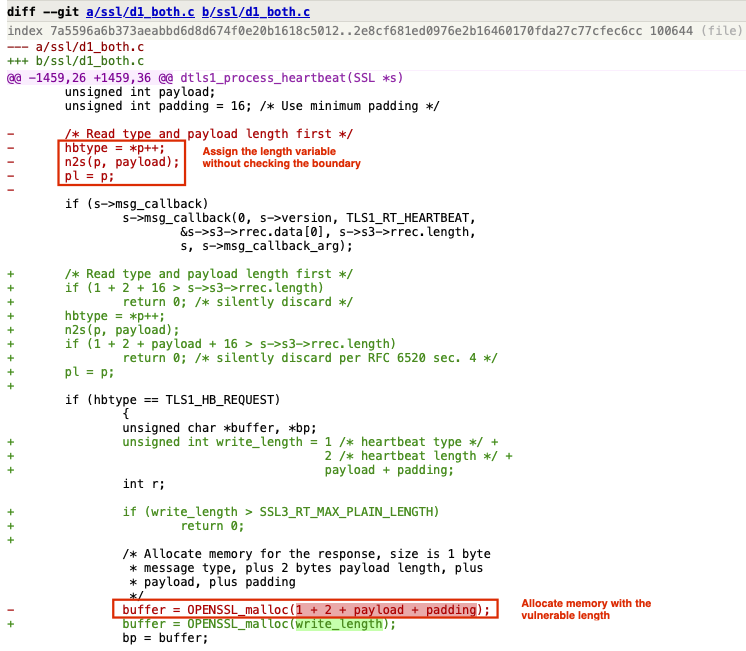}
\caption{Part of CVE-2014-0160 (Heartbleed) patch}
\label{fig:motivating-example-heartbleed-fix}
\end{figure}

\underline{Example 1:} \textit{Heartbleed} (CVE-2014-0160)\footnote{\url{https://nvd.nist.gov/vuln/detail/CVE-2014-0160}} in OpenSSL is a data leak \vulnerability{y} that occured due to improper input validation (CWE-20).\footnote{\url{https://cwe.mitre.org/data/definitions/20.html}} Heartbleed is one of the famous security incidents in OpenSSL that affected a large number of servers between 2012 and 2014. An OpenSSL client can send a \textit{Heartbeat} message to monitor the availability of a server. The server should return the received message to the client. However, in the vulnerable version (i.e., the version with the improper input validation), OpenSSL responds with the data of any length that the client specifies. 
Hence, sensitive information on the server can be obtained by the client. 
\reviewersComment{R3.2}
\reviewersModification{The \weakness{es} that caused such unexpected behavior is the improper input validation of the length parameter \texttt{write\_length}.}
As shown in the fixing patch of the \textit{Heartbleed} issue (see Figure \ref{fig:motivating-example-heartbleed-fix}), the vulnerable code was fixed by ensuring that the entered length is valid. A study by~\citet{Durumeric2014TheHeartbleed} suggested that if the reviewers had identified the improper input validation in the vulnerable version of OpenSSL, this \vulnerability{y} could have been prevented early.

\begin{figure}[h!]
  \includegraphics[width=\textwidth]{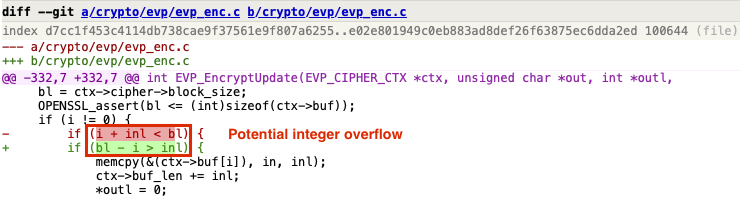}
\caption{Patch for CVE-2016-2106 - Fixing potential integer overflow expression in IF statement}
\label{fig:motivating-example-numeric-error-fix}
\end{figure}

\underline{Example 2:} The Integer Overflow or
Wraparound weakness can trigger heap memory corruption in a vulnerable version of OpenSSL, leading to a denial of service error (CVE-2016-2106\footnote{\url{https://nvd.nist.gov/vuln/detail/CVE-2016-2106}}). 
The cause of this \vulnerability{y} is the integer overflow (CWE-190).\footnote{\url{https://cwe.mitre.org/data/definitions/190.html}} The \textit{Integer Overflow or Wraparound} weakness is a numeric error where an integer value becomes larger than the maximum size of the associated data type, forcing the system to mistakenly wrap the value around and causing the denial-of-services error. 
\reviewersComment{R3.2}
\reviewersModification{
It can be seen in Figure \ref{fig:motivating-example-numeric-error-fix} that the original condition (\texttt{i + inl < b}) is prone to the integer overflow because the output of the left-hand-side operation can exceed the range of the implicit variable type.
This particular weakness was fixed by a patch that adjusted the expression by subtracting two integers, instead of adding them. } 
Hence, in this case, the denial-of-service vulnerability from heap memory corruption was caused by the integer overflow or wraparound weakness.

\begin{figure}[h]
  \includegraphics[width=\textwidth]{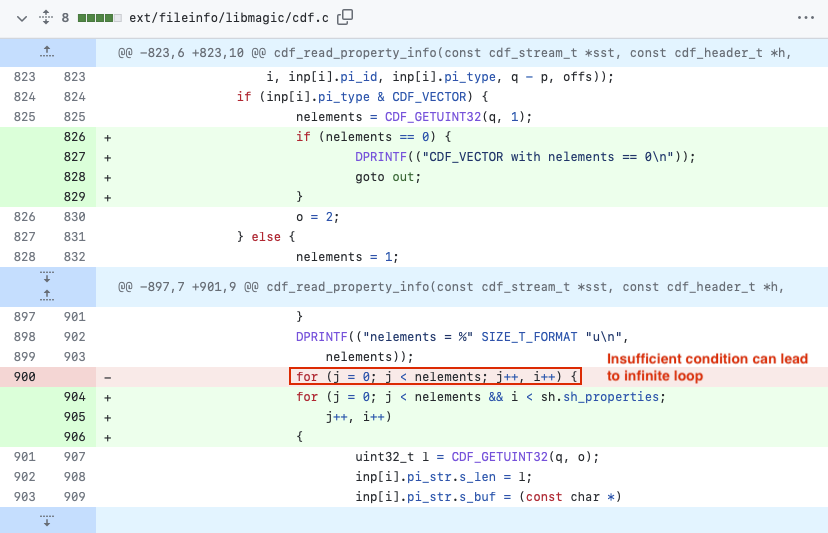}
\caption{Patch for CVE-2014-0238 - Fixing insufficient exit condition that may cause infinite loop}
\label{fig:motivating-example-business-logic-fix}
\end{figure}

\underline{Example 3:} The Potential Infinite Loop (CVE-2014-0238\footnote{\url{https://nvd.nist.gov/vuln/detail/CVE-2014-0238}}) is a denial-of-service \vulnerability{y} in PHP that was caused by a weakness of type \textit{Allocation of Resources Without Limits or Throttling} (CWE-770)\footnote{\url{https://cwe.mitre.org/data/definitions/770.html}}, which can also be considered as a \textit{Business Logic Errors} (CWE-840) weakness.\footnote{\url{https://cwe.mitre.org/data/definitions/840.html}} 
The vulnerable version of PHP can make the system unresponsive to requests when the attacker enters the lengthy input into a function that executes multiple for-loops.
\reviewersComment{R3.2}
\reviewersModification{As seen in Figure \ref{fig:motivating-example-business-logic-fix}, one of the loop variables (\texttt{i}) is incremented without being checked in the control condition.}
It was fixed by adding the missing exit condition to the loop when the counter reached the maximum size.

\begin{figure}[h!]
  \includegraphics[width=\textwidth]{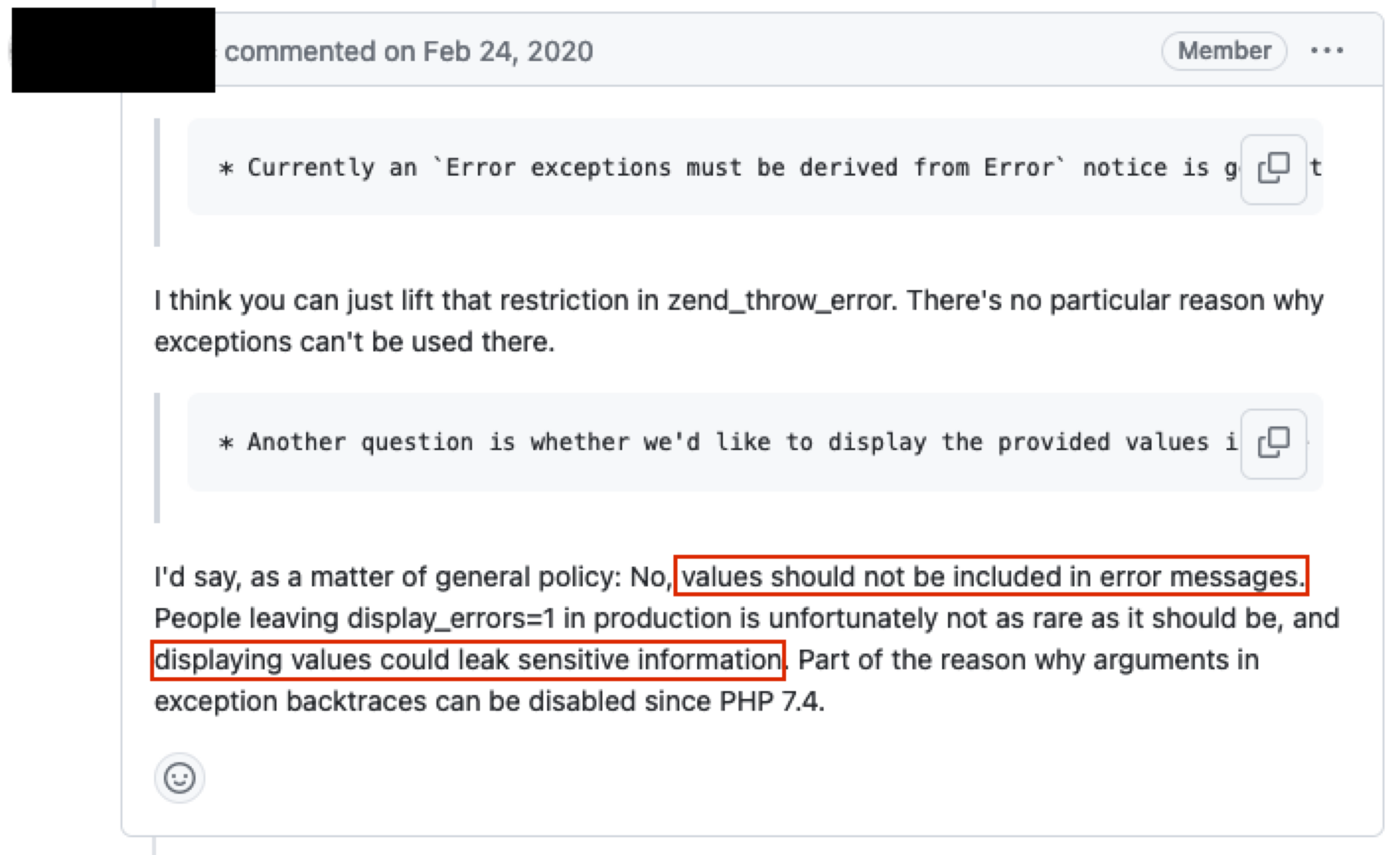}
\caption{Reviewer identified \weakness{} regarding the exposing of internal values in the error messages, which is relevant to Information Management Errors (CWE-199).}
\label{fig:motivating-example-weakness-found-in-review}
\end{figure}

These examples demonstrate that \weakness{es} can contribute to \securityIssue{s}. Since code reviews focus on identifying coding issues in source code, identifying coding weaknesses that can lead to various exploitable \vulnerability{ies} and \securityIssue{s} would be beneficial to code review practices.
\reviewersComment{R3.1}
\reviewersModification{
Refer to Figure~\ref{fig:motivating-example-weakness-found-in-review} for a real-world example of a \weakness{} i.e., exposing the internal values in error messages, relevant to \textit{Information Management Errors (CWE-199)}, identified during a code review.\footnote{\url{https://github.com/php/php-src/pull/5197\#issuecomment-590092307}}}

However, little has been investigated on how often the reviewers can identify \weakness{es} that link to security issues during the code reviews, what kinds of security concerns are raised, and how they are being handled or responded to. 
This insight would help practitioners improve their code review practice and equip developers with a secure code review mindset that is more compatible with their technical expertise.

\section{Case Study Design}
\label{design}
In this section, we outline our study by describing the research questions, the selection processes of studied subjects and \weakness{} taxonomy, the data collection method, and our analysis approach to answer our research questions.

As comments related to \securityIssue{s} are sparse in code reviews~\citep{DiBiase2016AChromium}, it can be challenging to synthesize an insightful result from diverse data sources. To overcome this problem, we followed the case study method~\citep{Perry2004CaseEngineers} to explore the real-world security concerns in selected software projects that are prone to \securityIssue{s}. We chose OpenSSL and PHP for our case study based on the \knownV{} vulnerabilities in the past and the code review activities.

\subsection{Research Questions}
To understand the potential benefits of considering coding weaknesses during code reviews for early prevention of
software \securityIssue{s}, we formulate the following research questions.

\inlineheading{RQ1: \rqOne}

\noindent\underline{Motivation}: Code review is an approach that many projects use to identify and eliminate defects early before integrating the new code into the codebase. However, \cite{Braz2022SoftwarePerspective} raised a concern that identifying \securityIssue{s} during the code review process can be challenging for developers because of the lack of security knowledge and awareness. 
As \cite{Bojanova2023BugVulnerabilities} showed that the chain of \weakness{es} can be the root cause of \securityIssue{}, we hypothesize that code reviews could identify such \weakness{es} which have simpler coding patterns and require less security knowledge.  
Prior studies have yet to investigate the security concerns that reviewers could raise from the \weakness{} perspective. Since \weakness{es} are more visible to developers than \securityIssue{s}, the understanding of these security concerns may provide guidance to improve \securityIssue{} identification in code reviews.

\vspace{1em}
\inlineheading{RQ2: \rqTwo}

\noindent\underline{Motivation}:
While RQ1 helps us to better understand what kind of \weakness{es} can be raised during the code review process, it is still unclear whether current code review practices have focused on \weakness{es} related to the real \vulnerability{ies} that were \knownV{} in the past. 
Thus, we set out RQ2 to examine the alignment of vulnerabilities that the systems had and the raised \weakness{} comments.
The findings will highlight the types of \weakness{es} that may not be sufficiently discussed in the code reviews. This understanding could increase the reviewer's awareness of the less frequently identified \weakness{es}.

\vspace{1em}
\inlineheading{RQ3: \rqThree}

\noindent\underline{Motivation}: Developers can respond to the raised \securityConcern{s} (i.e., \weakness{} raised during code reviews) in various ways in order to address the reviewer's comments and get the code accepted. \cite{Kononenko2016CodeIt} found that developers consider security concerns in reviewers' comments before modifying the code. In contrast, \cite{Lenarduzzi2021DoesStudy} reported that security defects do not influence the acceptance decision of the proposed code changes. However, little is known when it comes to the security concerns from \weakness{} comments. 
An extended understanding of how developers respond to these \securityConcern{s} could shed more light on the remaining challenges of current secure code review practice.
Hence, we set out to explore how the developers handle security concerns from \weakness{es}.

\subsection{Studied Projects}
\reviewersComment{R2.4}
\reviewersModification{
We aimed to conduct a case study of the code review process of software systems that are prone to \securityIssue{s}. Therefore, to select suitable projects, we considered the following criteria: 
}
\begin{enumerate}
    \item \reviewersModification{\textbf{Use C or C++ as the major programming languages}---Unlike other programming languages, 
    users of the C and C++ programming languages can access and manipulate lower-level environments \citep{Turner2014SecurityRuby} that are susceptible to security issues.}
    \item \reviewersModification{\textbf{Actively performing code reviews}---Quality assurance practices such as peer code reviews can improve the quality of code and reduce defects in the code base~\citep{Bacchelli2013ExpectationsReview, Beller2014ModernFix}.}
    \item \reviewersModification{\textbf{An accessible code review history}---To enable the data extraction, the complete code review data must be publicly accessible.}
\end{enumerate}

To find subjects prone to \securityIssue{s}, we obtained a list of C and C++ software systems that have records of publicly reported \vulnerability{ies} from the works of \cite{Hazimeh2020Magma:Benchmark} and \cite{Lipp2022AnDetection}.
We obtained nine software projects: \textit{Libpng}\footnote{\url{https://libpng.sourceforge.io}}, \textit{Libtiff}\footnote{\url{http://www.libtiff.org}}, \textit{Libxml2}\footnote{\url{https://gitlab.gnome.org/GNOME/libxml2}}, \textit{OpenSSL}\footnote{\url{https://www.openssl.org}}, \textit{PHP}\footnote{\url{https://www.php.net}}, \textit{Poppler}\footnote{\url{https://poppler.freedesktop.org}}, \textit{SQLite3}\footnote{\url{https://www.sqlite.org}}, \textit{Binutils}\footnote{\url{https://www.gnu.org/software/binutils/}}, and \textit{FFmpeg}\footnote{\url{https://ffmpeg.org}}. 

We first checked if the code review history of the project is publicly available as it is essential for our analysis. We also carefully checked whether the project regularly performs code reviews for every new code change by examining the project's code repository (e.g., GitHub and Gitlab), documents on the project's websites, and code review history in other sources (e.g., mailing list). 
\reviewersComment{R2.7}
\reviewersModification{
We found that \textit{OpenSSL} and \textit{PHP} established a public contribution policy\footnote{
OpenSSL-\url{https://github.com/openssl/openssl/blob/master/CONTRIBUTING.md}; PHP-\url{https://github.com/php/php-src/blob/master/CONTRIBUTING.md}} that requires
the developers to create GitHub pull requests and address any reviewer's comments for all public code submissions}; \textit{Libpng}, \textit{Libtiff}, and \textit{Libxml2} have relatively small code review history (less than 350 proposed code changes that received code review comments); \textit{SQLite3} employs a private code review process; \textit{Binutils} and \textit{FFmpeg} perform code review on mailing lists, which complicates the process of extracting code review information; and \textit{Poppler} has a large amount of code integration without reviewers' comments.

\reversemarginpar\reviewersComment{R1.1, R2.4}
\reviewersModification{Therefore, OpenSSL and PHP were the remaining projects that met our criteria. OpenSSL is a popular encryption library for secure communication over the Internet, and PHP is a widely used web scripting language. 
In terms of open-source project characteristics, OpenSSL receives more than 13k pull requests from nearly 900 active developers and gains over 23.8k stars on GitHub, while PHP, receives over 10k pull requests from over 900 active developers and gains approximately 37k stars on the same platform.
In addition to a remarkable number of active developers, both projects are important to the software development communities because numerous software rely on, or implement them.
As of July 2022, the number of CVE vulnerabilities reported was 215 and 662 for OpenSSL\footnote{\url{https://www.cvedetails.com/product/383/Openssl-Openssl.html}} and PHP\footnote{\url{https://www.cvedetails.com/product/128/PHP-PHP.html}}, respectively. Both projects regularly perform code reviews on GitHub\footnote{\url{https://github.com/openssl/openssl} and \url{https://github.com/php/php-src}}, where new proposed code changes are submitted as pull requests.}

\subsection{Data Collection}
\normalmarginpar\reviewersComment{R3.3}
\reviewersModification{
As we aimed to identify the types of \weakness{es} raised by reviewers and the handling process of these concerns, we needed to analyze the code review history, especially the review discussions. Therefore, we first downloaded the code review histories of the studied projects using the GitHub REST API.\footnote{\url{https://docs.github.com/en/rest}} 
We downloaded the pull requests and all comments on the pull requests. 
We accessed and retrieved all the code review historical data of both projects at the end of June 2022. 
The earliest pull requests that we downloaded from OpenSSL and PHP were created in September 2013 and July 2011, respectively.
}

\reviewersModification{
Then, we selected pull requests that are (1) closed, (2) proposed to upstream, i.e., the main branch, (3) comprised of at least one C or C++ file, and (4) have received at least one comment from a human reviewer.\footnote{
\reviewersModification{Comments that (1) matched the bot's characteristics~\citep{Golzadeh2020BotSimilarity} i.e., producing only comments that sustain the similar patterns and (2) were made by GitHub users of type \textit{Bot} were excluded.}}
}
\reversemarginpar\reviewersComment{R1.2}
\reviewersModification{
It should be noted that we considered the code review comments from (1) the main discussion of a pull request and (2) the code level because reviewers may discuss \weakness{es} in both levels.}
Table \ref{table:dataset_size} shows the number of downloaded and selected pull requests, as well as the number of comments on the selected pull requests.

\begin{table}[h]
\leftalignedcaption{Number of pull requests and code review comments}
\centering
\small
\begin{tabularx}{\linewidth}{
            >{\hsize=1\hsize}X
            >{\hsize=1\hsize}X
            >{\hsize=1\hsize}X
            >{\hsize=1\hsize}X
          }
\toprule
\textbf{Subject} & \textbf{Downloaded Pull Requests} &\textbf{Selected Pull Requests} & \textbf{Total Number of Comments}\\
\hline
 \midrule
 OpenSSL & 9,752 & 6,235 & 101,694 \\
 PHP & 5,718 & 3,497 & 33,866 \\ 
 \hline
\textbf{Total} & 15,470 & 9,732 & 135,560 \\
 \bottomrule
\end{tabularx}
\newline
\label{table:dataset_size}
\end{table}

\subsection{Coding Weakness Taxonomy}
Our objective was to classify security concerns in code review comments from the perspective of development flaws and not from the perspective of an attacker, which has been done in previous work \citep{Paul2021WhyProject, Bosu2014IdentifyingStudy, Bosu2013PeerEvaluation, Edmundson2013AnReview, DiBiase2016AChromium}. 
We aim to adopt the taxonomy that is generally applicable to the code reviews in any software system.\reviewersComment{R2.5}
\reviewersModification{
The selected taxonomy covers the realistic \securityConcern{s} from typical reviewers who may have fundamental software development expertise, but limited security knowledge and awareness~\citep{Braz2022SoftwarePerspective}.
}
Therefore, we selected an existing taxonomy based on the following criteria:

\begin{itemize}
    \item Covers the diverse \weakness{es} that are not restricted by the types of well-known \securityIssue{s}
    \item Provides detailed descriptions and examples for the ease of the annotation process and the future applicability for practitioners
    \item Focuses on the business and application logic that can be addressed in code rather than low-level aspects such as network or hardware 
    \item Unattached to a specific technology, programming language, or platform
\end{itemize}

We considered four \weakness{} taxonomies from industrial guidelines and standards: (1) Common Weakness Enumeration 699 (\textit{CWE-699}) \citep{CWE699}, (2) Common Weakness Enumeration 1000 \textit{CWE-1000} \citep{CWE1000}, (3) \textit{OWASP Code Review Guide} \citep{OwaspGuide}, and (4) OWASP Application Security Verification Standard (\textit{OWASP ASVS}) \citep{OWASPStandard}. 
Taking into account these four taxonomies based on our criteria, we opted to use CWE-699 in our study for the following reasons. CWE-699 covers a large number of \weakness{es} that can be linked to significant \securityIssue{s}. 
\reviewersComment{R2.5}
\reviewersModification{In particular, CWE-699 has 40 categories containing more than 400 weaknesses that may not expose the security implications (hence, do not require deep security knowledge to identify) but still lead to vulnerabilities and can be introduced during software implementation.}
On the contrary, the other taxonomies have a relatively smaller number of weaknesses (e.g., 6 categories~\citep{Croft2022AnLanguages}). 
Although CWE-1000 has a higher number of weaknesses, it includes weaknesses in other layers, e.g., hardware or platform, that can be outside of the code review context. CWE-699 provides an extended description of each of its weaknesses, while the OWASP Code Review Guide, which provides a checklist of nine critical vulnerabilities, does not provide a clear description of the suggested vulnerabilities that reviewers can utilize. As CWE-699 focuses on software implementation, its weaknesses are to some extent generalizable to any technology. On the other hand, OWASP ASVS which provides 14 categories of security requirements is more specific to web applications.
The simplified definitions of security weaknesses are shown in Table~\ref{table:cwe699-simplified}.

\begin{small}   
\begin{longtable}[l]{p{.3\textwidth} | p{.7\textwidth} } 
\leftalignedcaption{A list of \weakness{es} of the CWE-699 taxonomy }\\
\hline
    \textbf{Category Name} & \textbf{Simplified Description} \\
\hline
\endhead
\hline \multicolumn{2}{r}{\textit{Continued on next page}} \\
\endfoot
\hline
\endlastfoot

    API / Function Errors \newline (CWE-1228) &	Use of potentially harmful, inconsistent, or obsolete functions (built-in or API) or exposing low-level functionality in a function \\
    \hline
    Audit / Logging Errors \newline (CWE-1210) & Exposing excessive or sensitive data in the log, or logging insufficient security-related information that it cannot be used for forensic purposes \\
    \hline
    Authentication Errors \newline (CWE-1211) & Insufficient effort to verify requester’s identification either on a password-based or certification-based system, including bypassing authentication process \\
    \hline
    Authorization Errors \newline (CWE-1212) &	Insufficient check of actors' permitted capability before performing an operation on their behalf \\
    \hline
    Bad Coding Practices \newline (CWE-1006) &	Careless development or maintenance that leads to security issues e.g., active debug code \\
    \hline
    Behavioral Problems \newline (CWE-438) &	Unexpected behaviors by incorrectly written code e.g., infinite loop, missing bracket, misinterpretation of inputs, or a function that does not perform as documented \\
    \hline
    Business Logic Errors \newline (CWE-840) & Improper enforcement of usual features that allow an attacker to manipulate resources e.g., password reset without one-time-password \\
    \hline
    Communication Channel Errors \newline (CWE-417) &	Improper handling and verification of incoming and outgoing connections e.g., network connection or file access, which can lead to bypass attack \\
    \hline
    Complexity Issues \newline (CWE-1226) &	Various unnecessary complexities in software e.g., excessive parameters, large functions, unconditional branching, deep nesting, circular dependency, or in-loop update \\
    \hline
    Concurrency Issues \newline (CWE-557) &	Inconsistency caused by a race condition and incorrect synchronization on the particular resources \\
    \hline
    Credentials Management Errors \newline (CWE-255) &	Credential information such as password being stored, used, transported, exposed, or modified in the unsecured ways \\
    \hline
    Cryptographic Issues \newline (CWE-310) &	Insufficient use of cryptographic algorithms e.g., inappropriate key length, weak hash function, entropy generator, or signature verification \\
    \hline
    Key Management Errors \newline (CWE-320) &	Improper use of cryptographic keys e.g., hard-code key, using a key without checking for expiration, exchange key without verifying actors, or reusing nonce in encryption \\
    \hline
    Data Integrity Issues \newline (CWE-1214) &	Insufficient or negligence to verify or signify properties of data e.g., data source, data type, or checksum of incoming and outgoing data \\
    \hline
    Data Processing Errors \newline (CWE-19) &	Improper manipulation and handling such as parsing, extracting, and comparing of received data in terms of format, structure, and length \\
    \hline
    Data Neutralization Issues \newline (CWE-137) &	Lack of effort to prevent code injection and cross-site scripting from incoming data; and to enforce improper encoding and escaping characters for log and format-sensitive output such as CSV \\
    \hline
    Documentation Issues \newline (CWE-1225) &	Missing security-critical information in software document and inconsistency between documentation and actual implementation \\
    \hline
    File Handling Issues \newline (CWE-1219) &	Insecure and uncontrolled file path traversal or manipulation, including the use of insecure temporary file \\
    \hline
    Encapsulation Issues \newline (CWE-1227) &	Incorrect inheritance of program elements that may access restricted data or methods, not isolating system-dependent functions, or parent class refers children's properties \\
    \hline
    Error Conditions, Return Values, Status Codes \newline (CWE-389) &	Unwanted actions around return and exception i.e., sensitive return value, not checking returned value, not handling exception, unknown status code, or silently suppressing errors \\
    \hline
    Expression Issues \newline (CWE-569) &	Wrong expression in condition statements  i.e., always true or false boolean value, incorrect operators, comparing pointers, or incorrect expression precedence \\
    \hline
    Handler Errors \newline (CWE-429) &	Missing or improperly implementing the handling functions of an object e.g., intertwining handler functions, or calling non-reentrant functions in the handler function \\
    \hline
    Information Management Errors \newline (CWE-199) &	Leak sensitive information or internal states to unexpected actors via any channels \\
    \hline
    Initialization and Cleanup Errors \newline (CWE-452) &	Improper initialization and exit routine e.g., using excessive hard-coded data for initialization or not cleaning up the referenced memory space \\
    \hline
    Data Validation Issues \newline (CWE-1215) &	Insufficient validation of non-text data that can lead to security flaws e.g., array index, loop condition, or XML data; including early validation and improper restrictions (whitelist and blacklist) to validate against \\
    \hline
    Lockout Mechanism Errors \newline (CWE-1216) &	Overly restrictive of account lockout mechanism after a certain number of failed attempts \\
    \hline
    Memory Buffer Errors \newline (CWE-1218) &	Insufficient handling of a buffer that leads to security problems i.e., copy external data without checking the size, incorrect size calculation, buffer overflow-underflow, out-of-bound \\
    \hline
    Numeric Errors \newline (CWE-189) &	Improper calculation, conversion, comparison of numbers e.g., integer overflow or underflow, wraparound, truncation, off-by-one, endian, or division by zero; which could yield security problems \\
    \hline
    Permission Issues \newline (CWE-275) &	Improper assignment of permissions on file, object, and function during installation or inheritance within a program \\
    \hline
    Pointer Issues \newline (CWE-465) &	Improper handling of pointer e.g., accessing, returning, or scaling pointer outside the range, dereferencing the invalid pointer, using fixed pointer address, or accessing and releasing wrong pointer \\
    \hline
    Privilege Issues \newline (CWE-265) &	Assigning improper system privileges that violate least privilege principle e.g., non-terminated \textit{chroot jail}, including incorrectly handling or revoking insufficient privileges  \\
    \hline
    Random Number Issues \newline (CWE-1213) &	Use predictable algorithm or insufficient entropy for critical random number generator \\
    \hline
    Resource Locking Problems \newline (CWE-411) &	Improper control of resource locking e.g., not checking for prior lock, improper degree of resource locking, or deadlock \\
    \hline
    Resource Management Errors \newline (CWE-399) &	Improper organization of system in general sense e.g., allowing external influence to access file, class, or code; not releasing resource after use, or using after release; duplicating resource identifier \\
    \hline
    Signal Errors \newline (CWE-387) &	Improper handling of signal e.g., race condition triggered by handlers, allowing active handler during sensitive operations, not handling with asynchronous-safe functionality, handling multiple signals in single function \\
    \hline
    State Issues \newline (CWE-371) &	Allow the program to enter the unexpected state by multiple causes e.g., letting external actor control system setting, passing mutable objects to an untrusted actor, calling a non-reentrant function that calls other non-reentrant function in nested call \\
    \hline
    String Errors \newline (CWE-133) &	Use external format string, incorrectly calculate multi-byte string length, or use wrong string comparison method \\
    \hline
    Type Errors \newline (CWE-136) &	Access or converse resources with incompatible types i.e., type confusion error \\
    \hline
    User Interface Security Issues \newline (CWE-355) &	Storing sensitive information in the UI layer, having an insufficient warning for unsafe or dangerous actions, having an unsupported or obsolete function in the user interface, allowing Clickjacking \\
    \hline
    User Session Errors \newline (CWE-1217) &	Leak data element in wrong session or permitting unexpired session \\
    
\label{table:cwe699-simplified}
\end{longtable}
\end{small}



\subsection{Study Overview}

\begin{figure*}[ht!]    
    \centering
    \resizebox{\linewidth}{!}{\includegraphics{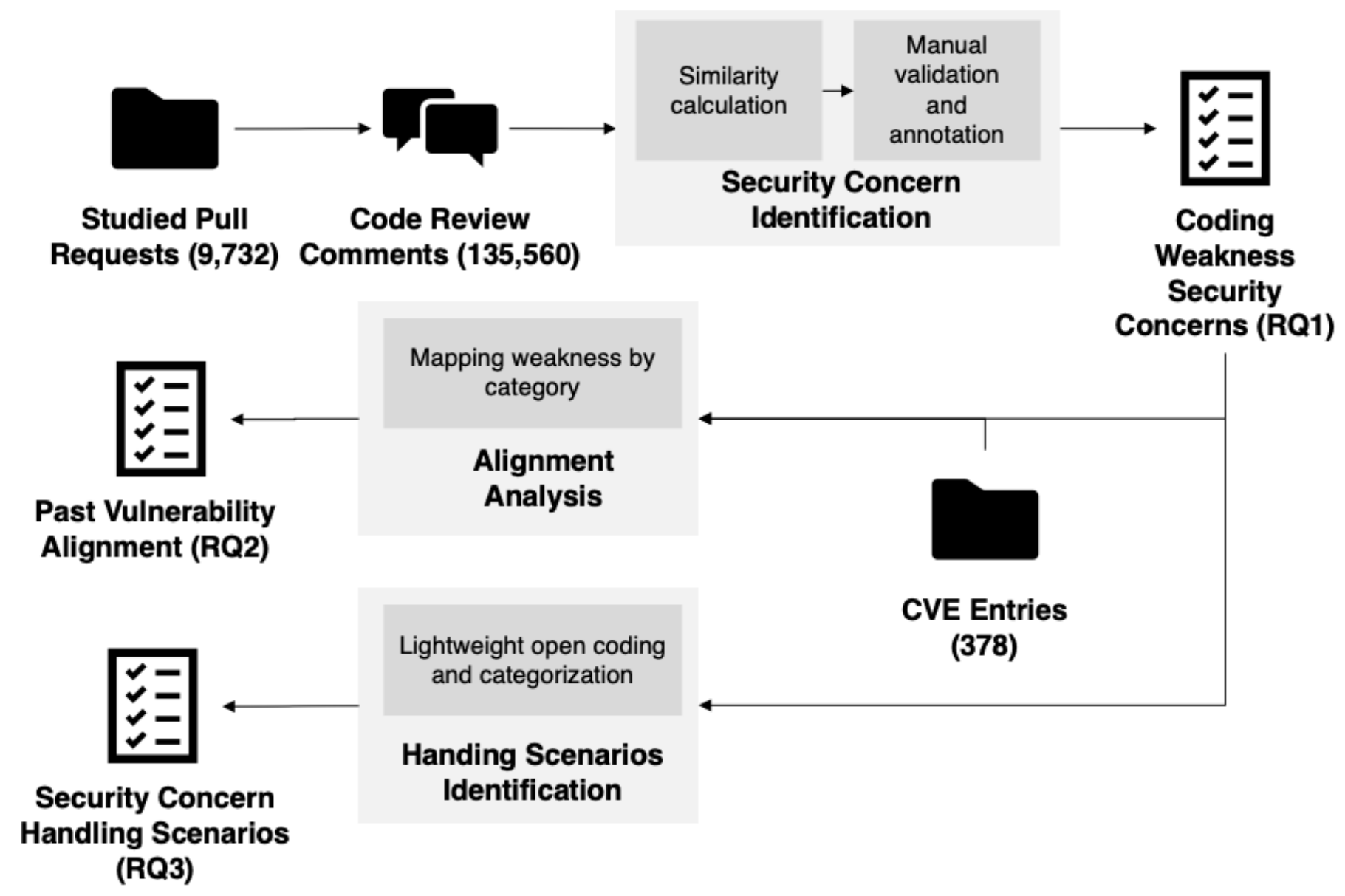}}
    \leftalignedcaption{Overall case study workflow}
    \label{fig:overall-workflow-diagram}
\end{figure*}

Figure \ref{fig:overall-workflow-diagram} shows an overview of our study. To explore the \weakness{es} raised by the reviewers (RQ1), we use a semi-automated approach to identify comments that raised \weakness{es} in a pull request. In particular, we apply text analysis techniques to sort comments that have high semantic similarity with the descriptions of \weakness{es} in the CWE-699 taxonomy. Then, we manually validate and annotate the types of security weaknesses based on the CWE-699 taxonomy.  
The \securityConcern{s} found in RQ1 will be further qualitatively analyzed for RQ2 and RQ3, along with their related pull request information and \knownV{} vulnerabilities.
For the alignment assessment with \knownV{} \vulnerability{ies} (RQ2), we analyze the \vulnerability{ies} of the studied systems that were reported in the past against the \securityConcern{s} raised during the code reviews.
For RQ3, we qualitatively analyze the related code review comments and corresponding code changes within the pull requests to investigate the handling scenario of the raised \securityConcern{s}. 

\reviewersModification{
\reviewersComment{R2.1, R2.10}
The semi-automated approach (RQ1) provided dual benefits. 
Firstly, it addressed the challenges in identifying relevant code review comments, overcoming limitations faced by previous studies~\citep{Bosu2014IdentifyingStudy, Paul2021WhyProject, Alfadel2023EmpiricalPackages} that can only identify the limited set of issues in code review comments by using the keyword-based approach. 
Thus, it enabled the discovery of more comments related to \weakness{es}. Secondly, it mitigated the manual effort required for validating comments associated with each type of \weakness{}. However, manual validation and annotation remain essential to eliminate false positives beyond the capacity of the automated process.
}


We explain our study approaches in the following sections: Section~\ref{automated_concern_identification} describes the security concern identification approach for RQ1,  Section~\ref{concern_manual_validation} describes the handling scenarios identification approach for RQ3, and Section \ref{mapping_past_vulnerability} describes the \knownV{} vulnerabilities alignment analysis approach for RQ2.

\subsection{Security Concern Identification Approach (RQ1)}
\label{security_concern_identification_approach}
Due to a large number of code review comments (i.e., 135K comments in our dataset; see Table \ref{table:dataset_size}), it is not feasible for us to manually identify security comments related to \weakness{es} (so-called, \securityConcern{s}). 
\reviewersComment{R2.6}
\reviewersModification{
We opted to use an automated approach to support our \securityConcern{} identification process.
While prior works~\citep{Bosu2014IdentifyingStudy, Paul2021WhyProject, Alfadel2023EmpiricalPackages} used the keyword-based approach to identify the comments that mentioned \vulnerability{ies}, their pre-defined keyword lists are limited to the specific types of security issues which do not cover all categories in CWE-699. 
}
Indeed, the keyword-based approach does not find the code review comments that contain other \weakness{es} that are not included in the keyword lists.
\reviewersModification{
Therefore, we resorted to employing a semi-automated approach that combines the textual analysis technique using pre-trained word-embedding model to rank comments that are semantically related to \weakness{es} and human evaluation. 
}

Our approach comprises two steps: 1) calculating semantic similarity with the descriptions of \weakness{es} and 2) manually validating and annotating the types of security weaknesses based on the CWE-699 taxonomy.
We now provide the details of each step below.

\subsubsection{Semantic Similarity Calculation}
\label{automated_concern_identification}
Our underlying intuition is that code review comments that are semantically similar to the descriptions of the \weakness{es} could be considered as \weakness{} security comments. Therefore, we use the cosine similarity score to measure how a code review comment is similar to a \weakness{}. Cosine similarity is a widely used technique for document similarity calculation because it does not consider the length of documents. The cosine similarity score can be calculated using the following formula.

\begin{equation}
\text{Comment-Weakness Cosine Similarity} = \frac{\mathbf{C} \cdot \mathbf{W}}{\|\mathbf{C}\| \|\mathbf{W}\|}
\end{equation}

Where $\mathbf{C}$ and $\mathbf{W}$ are the vectors of the code review comment and description of a \weakness, respectively.
Prior to creating the vectors, we prepare the comments and the description of each of the 40 categories in CWE-699 by removing hyperlinks, stopwords, numbers, and non-alphanumeric characters. 
We also apply \textit{SnowballStemmer} \citep{Snowball} to obtain the common form of each word.
Then, we generate the vector of each code review comment and 40 vectors of the descriptions of \weakness{es} by using the Gensim library~\citep{Rehurek2011GensimpythonModelling} and calculate the similarity scores.  
Therefore, each code review comment will receive 40 similarity scores.
A high similarity score for a particular category indicates that the comment is more likely related to that category.

The results of cosine similarity calculation strongly depend on the vector representation of the documents. In order to select the vector representation that yields the best results, we explore two vector representation techniques.

\begin{itemize}

    \item \textbf{Term Frequency - Inverse Document Frequency (TF-IDF): }
     We measure the cosine similarity between each code review comment against the full description of each of the CWE-699 weakness categories. We used Term Frequency - Inverse Document Frequency (TF-IDF) vectors \citep{Tata2007EstimatingPredicates} to represent the code review comment and the description of the weakness. A TF-IDF vector represents the significance of every word in a document (i.e., code review comment and the weakness description). The Term Frequency (TF) is the number of times that a word appears in a document over the total number of words in the document, and the Inverse Document Frequency (IDF) is the logarithm of the ratio between the number of all documents and the number of documents that contain the word. The TF-IDF score for each word is the multiplication of the TF value and IDF value. The TF-IDF vector contains the TF-IDF scores of all words in a document.
    
    \item \textbf{Word Embedding:}
    Developers may use interchangeable terms during code review which may be different from the content of the weakness description. For example, \textit{login} may refer to \textit{authenticate}  in some contexts. Conventional vector representations such as TF-IDF may obstruct the expansive definition of a word. A soft cosine similarity technique that incorporates word embedding models with cosine similarity \citep{Ye2016FromEngineering, Sidorov2014SoftModel} can mitigate this limitation. Instead of using TF-IDF vectors, we use word embedding models to generate vectors that represent the code review comment and the full description of weaknesses. We explore a pre-trained word embedding model in the software engineering domain, namely \textit{SO Word2Vec} \citep{Efstathiou2018WordDomain}. 
    It should also be noted that we carefully check if the words are included in word-embedding model before the stemming process. This is to address the out-of-vocabulary (OOV) problem because we stem and retain the words that are not in the word-embedding model for similarity score calculation.
\end{itemize}

\subsubsection{Manual Validation \& Annotation of Security Concerns}
\label{concern_manual_validation}

\begin{figure}[h]
    \centering
    \resizebox{\linewidth}{!}{\includegraphics{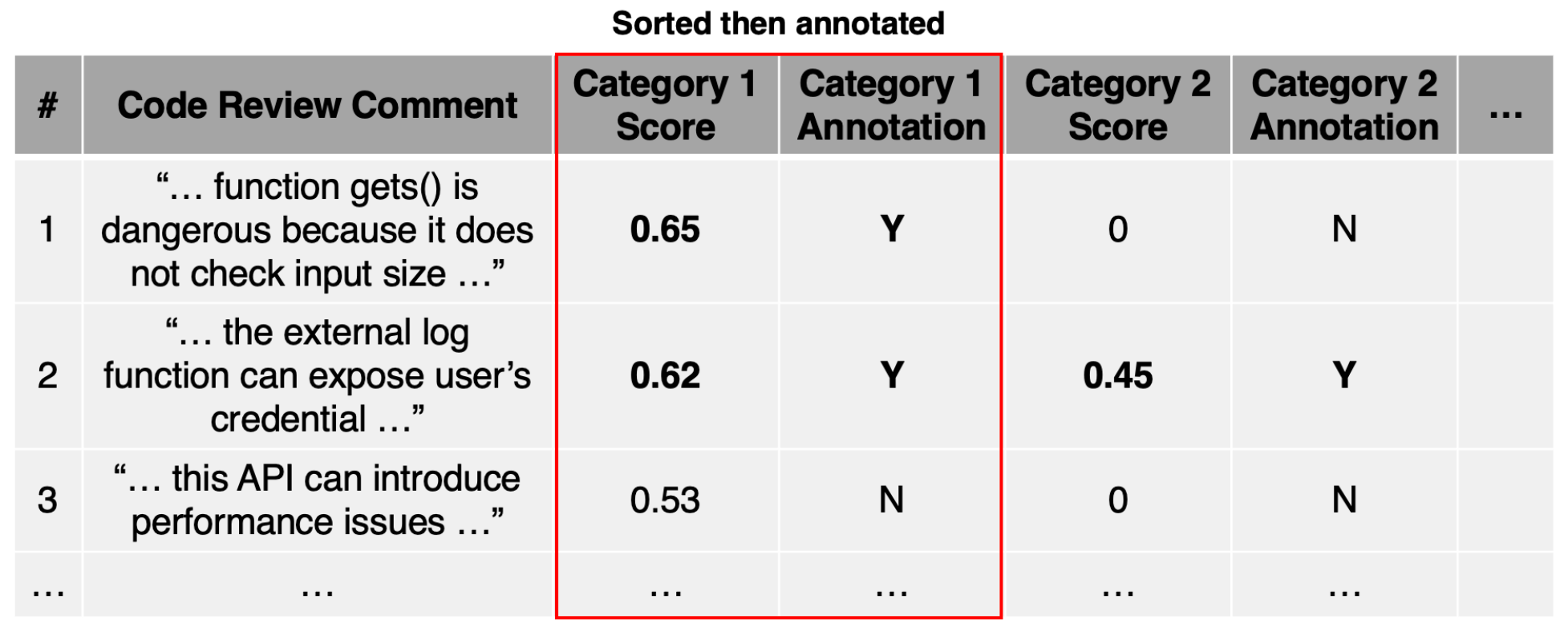}}
    \leftalignedcaption{An illustration of our manual validation and annotation approach.}
    \label{fig:annotation-process-example}
\end{figure}

\reviewersComment{R1.3}
\reviewersModification{Once we calculated semantic similarity scores for each code review comment against each of the \weakness{} descriptions in the CWE-699 taxonomy, we manually validated and annotated the code review comments into the CWE-699 taxonomy. In particular, we manually analyzed the comments from those with the highest similarity scores to those with lower scores. Specifically, for each category, the comments were sorted by similarity score in descending order before manually annotating the entire body of each comment to determine whether a security concern that is relevant to the \weakness{es} was raised.
Figure~\ref{fig:annotation-process-example} illustrates an example of our manual validation and annotation approach.}

\reviewersComment{R2.8}
\reviewersModification{
We performed manual validation and annotation in two rounds.}

\reviewersModification{
\textbf{First Round--Screening}
The first round aimed to preliminarily remove comments that are generic or irrelevant to \weakness{} (e.g., related to bookkeeping and code styling). 
For each CWE-699 category, we carefully read comments and determine if they are related to that category based on its description (see Table~\ref{table:cwe699-simplified}). Due to the large number of comments, it is not feasible to validate all the comments across the 40 categories (i.e., 40 $\times$ 135K in total). Thus, for each category,  we validate the comments until reaching the saturation point, i.e., 50 consecutive comments were identified as generic or irrelevant comments. 
In total, we validated 9,704 scores of 6,146 code review comments from both projects. Note that one comment can have high similarity scores in multiple categories. For example, Comment\#2 has high similarity scores in categories 1 and 2 (see Figure~\ref{fig:annotation-process-example}). Hence, such a comment will be read and validated more than once.  
This screening process in the first round was done by the first author (30 categories) and a third-year PhD student, who is experienced in manual analysis, (10 categories). 
}

\reviewersModification{
\textbf{Second Round--Validation}
The second round aimed to carefully identify security concerns from the comments that passed the first round.
In particular, we determined whether the comments raised legitimate security concerns that are relevant to the \weakness{es}. Specifically, a comment was determined to contain a legitimate security concern if it was relevant to one of the CWE-699 categories and met at least one of the following two conditions:
}

\begin{itemize}
    \item \textbf{Reviewer purposefully remarked security consequence(s).} For example, the reviewer commented that \textit{"Are these implementations safe against charset based attacks?."}\footnote{\url{https://github.com/php/php-src/pull/167\#issuecomment-7852838}} which can be considered to indicate a problem with the neutralization of the input data.
    \item \textbf{Reviewer expressed concern that can potentially point to a security issue, but did not explicitly mention security consequence(s)}. For example, the reviewer commented that \textit{"[...] login which always asks for name/password but returns 'no such user' or 'wrong password'  Nowadays they always return 'Bad login'"}\footnote{\url{https://github.com/openssl/openssl/pull/8300\#issuecomment-582041297}} which may indicate an improper authentication process.
\end{itemize}

When identifying concerns, it is possible that multiple comments in the same pull request indicated the same security concern. For example, two reviewers suggested that the developer verify a concurrency issue in the code.\footnote{\url{https://github.com/php/php-src/pull/5543\#issuecomment-627313863} and \url{https://github.com/php/php-src/pull/5543\#issuecomment-627313863}} In such a case, we consider these comments as a single security concern for that pull request. 
Similarly, a comment can also contain multiple legitimate concerns. For example, a reviewer raised concerns about the predictability of the random number generator algorithm\footnote{\url{https://github.com/openssl/openssl/pull/9423\#issuecomment-513757655}} which can be interpreted as both Random Number Issues and Cryptographic Issues concerns based on the CWE-699 categories.
In this case, we considered that this pull request has two concerns.

\reviewersComment{R2.8}
\reviewersModification{
The second round of annotation was done independently by the first author and the third author, who has ten years of experience in security testing.
To establish a clear understanding and ensure a consistent classification in the second round, the first author and the second author performed co-classification on a small set of comments to establish a common understanding of the annotation task.}
Both authors independently classified the remaining comments, and the inter-rater agreement was measured. It should be noted that the authors also assessed the code changes and the code review conversation to better understand the context when necessary. We used Cohen's Kappa \citep{Cohen1960AScales} to measure the inter-rater agreement. For comments with disagreement between the two authors, all authors met and discussed how to resolve the conflicts. The code review comments that were annotated from this process would be called \securityConcern{s} in the later steps.

\reviewersComment{R3.4}
\reviewersModification{
To ensure that the chance of missing the relevant comments is minimized, we also assessed the false negative rate on 400 randomly sampled unseen comments, which were left out after the saturation point in each category was reached, from each project.
We found very few unseen comments (two in OpenSSL and three in PHP) that can be considered security and weakness-related. 
In comparison to the ratio of positively identified comments from annotated data, we considered the false negative rate insignificant.
}

\subsection{Alignment Analysis of \knownVC{} Vulnerabilities (RQ2) }
\label{mapping_past_vulnerability}
To answer our RQ2, we further analyzed the \securityConcern{s} raised during the code reviews obtained from RQ1 against the \knownV{} \vulnerability{ies} of the studied systems that were reported in the past. 
To study the past \vulnerability{ies} in our studied projects, we used Common Vulnerabilities and Exposures (CVE) which is the collection of publicly reported vulnerabilities in software systems.  We downloaded the CVE entries of OpenSSL and PHP from the CVE mirror database\footnote{\url{https://www.cvedetails.com}}.  We collected the CVE entries that were reported within the same timeframe as the analyzed code reviews (2013-2022 in OpenSSL; 2011-2022 in PHP). 
We downloaded a total of 101 CVEs for OpenSSL and 277 CVEs for PHP. We then excluded the CVEs that were not assigned any CWEs since we would use the assigned CWE numbers to map into the CWE-699 taxonomy. In addition, we excluded the CVEs that have deprecated CWEs.
Finally, we studied 81 and 236 CVEs for OpenSSL and PHP, respectively.
Table~\ref{table:cve-mapping-statistics} shows the number of CVEs used in this study. 

\begin{table}[t]
\centering
\small
\leftalignedcaption{Number of CVEs in this study}
\begin{tabularx}{\linewidth}{
            >{\hsize=2.2\hsize}X
            >{\hsize=0.4\hsize}X
            >{\hsize=0.4\hsize}X
          }

\toprule
 \textbf{} & \textbf{OpenSSL} &\textbf{ PHP} \\ [0.5ex] 
 \midrule
    Downloaded CVEs & 101 & 277 \\
    \underline{Excluded}: & & \\
    \hspace{3mm} CVEs that are not assigned any CWEs & 17 & 40 \\    
    \hspace{3mm} CVEs that cannot be mapped due to CWE deprecation & 3 & 1 \\ 
    \hline
    \textbf{Studied CVEs} & \textbf{81} & \textbf{236} \\
 \bottomrule
\end{tabularx}
\newline
\label{table:cve-mapping-statistics}
\end{table}

To examine the alignment between the \securityConcern{s} raised during the code reviews and the \vulnerability{ies} of the studied systems that were reported in the past, we quantitatively analyzed the frequency of the concerns and the CVE entries across the 40 categories of the CWE-699 taxonomy. While each CVE entry has an assigned CWE number, it may not be directly associated with the CWE-699 categories.
A clear mapping between CVE entries and CWE-699 categories is not available because a CVE entry can be assigned with any CWE number which may not be under CWE-699 categories.
Therefore, we used the CWE hierarchical tree to systematically map the CVE entry and its assigned CWE number into the relevant CWE-699 categories. In particular, we considered the CVE entry to be relevant to the \texttt{category A} in the CWE-699 categories when
\begin{itemize}
    \item The assigned CWE has the same CWE as the \texttt{category A}; or
    \item The assigned CWE has a relationship (e.g., \texttt{PeerOf}, \texttt{CanFollow}, and \texttt{CanAlsoBe}) with CWE of \texttt{category A}; or
    \item The assigned CWE is the child of CWE of \texttt{category A}.
\end{itemize}

To illustrate the CWE mapping process, we provide an example for each condition as follows. For the first condition, CVE-2014-8275\footnote{\url{https://nvd.nist.gov/vuln/detail/cve-2014-8275}} has been assigned with CWE-310\footnote{\url{https://cwe.mitre.org/data/definitions/310.html}} which is the category Cryptographic Issues. 
For the second condition, CVE-2014-3670\footnote{\url{https://nvd.nist.gov/vuln/detail/CVE-2014-3670}} is relevant to category Numeric Errors (CWE-189)\footnote{\url{https://cwe.mitre.org/data/definitions/825.html}} because it has been assigned with CWE-119\footnote{\url{https://cwe.mitre.org/data/definitions/119.html}} that has the \texttt{CanFollow} relationship with CWE-128\footnote{\url{https://cwe.mitre.org/data/definitions/128.html}} which belongs to category CWE-189.
For the third condition, CVE-2017-12932\footnote{\url{https://nvd.nist.gov/vuln/detail/cve-2017-12932}} is considered as relevant to category Pointer Issues (CWE-465)\footnote{\url{https://cwe.mitre.org/data/definitions/465.html}} because it has been assigned with CWE-416\footnote{\url{https://cwe.mitre.org/data/definitions/416.html}} that is the child of CWE-825\footnote{\url{https://cwe.mitre.org/data/definitions/825.html}} which belongs to category CWE-465.

Note that an assigned CWE can be relevant to multiple CWE-699 categories. In that case, we classify that CVE entry as relevant to every related CWE-699 category. 
If the assigned CWEs of a CVE cannot be mapped based on the heuristic above, we manually map them into the relevant CWE-699 categories based on the descriptions. In particular, 29 (out of 81 for OpenSSL) and 81 (out of 236 for PHP) CVEs require a manual mapping of CWEs based on the descriptions.
Table~\ref{table:cve-cwe-manual-map} shows a list of CWEs that we manually mapped into the CWE-699 categories. 

\begin{table}[h]
\centering
\small
\leftalignedcaption{A manual mapping between assigned CWEs in CVEs and the CWE-699 taxonomy when the assigned CWEs have a direct relationship with the CWE-699 categories.}
\begin{tabularx}{\linewidth}{
            >{\hsize=0.25\hsize}X
            >{\hsize=1.5\hsize}X
            >{\hsize=1.75\hsize}X
            >{\hsize=0.5\hsize}X
          }
\toprule
\textbf{CWE} & \textbf{CWE Name} &\textbf{CWE-699 Category} & \textbf{Note}\\
\hline
 \midrule
    119	&	Improper Restriction of Operations within the Bounds of a Memory Buffer	&	Memory Buffer Errors (CWE-1218)	& \\
    200	&	Exposure of Sensitive Information to an Unauthorized Actor	&	Information Management Errors	(CWE-199)	& \\
    264	&	Permissions, Privileges, and Access Controls	&	Authentication Errors	(1211)	& \\
    287	&	Improper Authentication	&	Authentication Errors	(CWE-1211)	& \\
    326	&	Inadequate Encryption Strength	&	Cryptographic Issues	(CWE-310)	& \\
    362	&	Concurrent Execution using Shared Resource with Improper Synchronization ('Race Condition')	&	Concurrency Issues	(CWE-557)	& \\
    400	&	Uncontrolled Resource Consumption	&	Resource Management Errors	(CWE-399)	& \\
    674	&	Uncontrolled Recursion	&	Data Processing Errors	(CWE-19)	& \\
    16	&	Configuration	& -- & Deprecated CWE\\
    17	&	DEPRECATED	& -- & Deprecated CWE\\
 \bottomrule

\end{tabularx}
\newline
\label{table:cve-cwe-manual-map}
\end{table}

\subsection{Handling Process Identification (RQ3)}
\label{handling_process_identification}

To understand how developers handle security concerns (i.e., code review comments that mention \weakness{es} that could lead to \securityIssue{s}) (RQ3), we employed a lightweight coding method, similar to the approach used by \cite{Gousios2014AnModel}, to analyze what happens after security concerns are raised. We analyzed the security concerns that were manually identified in RQ1. 
    
Our aim was to identify the handling scenario based on the code review activities that occurred after security concerns were raised.
To do so, for each security concern, we read the discussion in the pull request, including the developer responses and the subsequent comments by the reviewer or other reviewers. In addition, similar to prior work \citep{Rahman2017PredictingExperience}, if the security concern points to particular lines of code, we checked whether the developer subsequently modified the associated lines of code to address the concern. Then, we summarized the observation for each security concern.
Finally, the summarized observations were sorted into groups on the basis of their thematic similarities and a handling scenario theme was defined for each group. 

\reviewersModification{
\reviewersComment{R2.8, R3.5}
The manual analysis in RQ3 was done by the first author and the second author. 
We analyzed the handling scenarios in three steps. 
In the first step, the first author summarized the discussion in a pull request into a brief note describing the handling of each security concern. 
In the second step, the first author reviewed the notes and categorized them into distinct groups based on thematic similarities. 
In the last step, the second author reviewed the groups before discussing the disagreed groups with the first author. 
Then, the first author refined the groups according to the mutual agreement.
Following refinement, the first author revisited the notes and ensured that they fit with the refined groups. 
We repeated the second and last step in multiple iterations until no further changes were needed (i.e., no new groups emerged, and all notes remained in the original groups). 
Finally, the second author manually validated 10\% of the security concerns to confirm the correctness of the annotated scenarios.
}

\section{Preliminary Analysis}
\label{preliminary_analysis}
\reviewersComment{R2.6}
\reviewersModification{In this section, we present two preliminary analyses to provide the logical ground for our main case study. 
The goal of the first preliminary analysis (PA1) is to examine whether reviewers tend to raise \weakness{es} related to \securityIssue{s} more frequently than explicitly discussing the \vulnerability{ies}.
The second analysis (PA2) aims to preliminarily evaluate the effectiveness of our semi-automated approach (see Section \ref{automated_concern_identification}) to calculate semantic similarity scores for the code comments that contain \weakness{es}.
}

\underline{Dataset}: We conducted the two preliminary analyses based on a sample dataset. We randomly sampled 400 code review comments from each of the studied projects (i.e., OpenSSL and PHP). This sample size should allow us to generalize conclusions with a confidence level of 95\% and a confidence interval of 5\%~\citep{Triola2009ElementaryStatistics}. 


\subsection{PA1: Prevalence of Coding Weakness Comments}
The motivating examples in Section~\ref{motivation} show that \weakness{es} can lead to \securityIssue{s}. 
Since code review focuses on identifying and mitigating issues in source code~\citep{Mantyla2009WhatReviews, Bacchelli2013ExpectationsReview}, it is possible that code review may be able to identify such \weakness{es}. 
To confirm this, we assess the degree to which the \weakness{es} are discussed in code reviews.
In particular, we analyze whether reviewers more frequently discussed \weakness{es} than \vulnerability{ies}.

\smallheading{Approach}
From the sampled dataset, we manually classified code review comments into three groups: 1) comments that mentioned a \weakness{}, 2) comments that explicitly mentioned a \vulnerability{y}, and 3) other comments that are not related to \weakness{es} and vulnerabilities.
We consider that a code review comment mentioned a \weakness{} when it is related to \weakness{es} listed in the CWE-699.
A code review comment is considered as mentioning a \vulnerability{y} when it is related to the types of exploitable vulnerabilities obtained from prior studies \citep{DiBiase2016AChromium, Paul2021WhyProject} i.e., Race Condition, Buffer and Integer Overflow, Improper Access, Cross Site-Scripting (XSS) and Cross-Site Request Forgery (CSRF), Denial of Service (DoS) and Crash, Information Leakage, Command and SQL Injection, Format String, Encryption, and common vulnerability keywords such as attack, bypass, back-door, breach, trojan, spyware, virus, ransom, malware, worm, and sniffer.
Note that one code review comment can be classified into multiple categories.
For example, \textit{a comment '[..] we ensure that when the `while` loop ends, there are always at least 2 more slots available in the \underline{output buffer without overrunning it} [..]'\footnote{\url{https://github.com/php/php-src/pull/8257}}} is related to a \vulnerability{y} (i.e., buffer overflow)  as well as a \weakness{} (Incorrect Calculation of Buffer Size (CWE-131)).
Hence, this comment is classified as mentioning \vulnerability{y} and \weakness{}.



\smallheading{Results}
Our preliminary result shows that \weakness{es} were raised more often than \vulnerability{ies} during the code review.
Table~\ref{table:weakness-prevalence} shows the number of code review comments that mentioned a \weakness{}, a \vulnerability{y}, and others. 
From 400 sampled code review comments for each studied project, we identified 67 comments related to \weakness{es} and 2 comments related to \vulnerability{ies} in PHP; and 84 comments related to \weakness{es} and 4 comments related to \vulnerability{ies} in OpenSSL. 
The amount of code review comments that mentioned \vulnerability{ies} align with the findings of~\cite{DiBiase2016AChromium} who found that 1\% of the code review comments identified \vulnerability{ies}.

Table~\ref{table:weakness-prevalence} shows that the number of comments that mentioned a \weakness{} is 21 - 33.5 times higher than the number of comments that mentioned a \vulnerability{y}.
In addition, we observed that reviewers sometimes point out a potential impact on the system when raising a \weakness{}.
For example, a reviewer commented that \textit{`[..] bad things will happen if the object gets freed before the \texttt{ctx} has finished using it [..]`}~\footnote{\url{https://github.com/openssl/openssl/pull/1372\#discussion_r73132165}} which is related to the 
`Use After Free' weakness CWE-416 and express a concern on its potential impact (i.e., "\textit{bad things will happen}").
These findings indicate that code review comments related to \weakness{es} are more prevalent than code review comments related to \vulnerability{ies}, highlighting that \weakness{es} which are the faults in software development can be identified and discussed during the code review process.

\begin{table}[t]
\centering
\small
\leftalignedcaption{Number of classified code review comments in the sampled dataset from each project (n=400). Note that one code review comment can be classified into multiple categoties.}
\begin{tabularx}{\linewidth}{
            >{\hsize=2\hsize}X
            >{\hsize=0.5\hsize}X
            >{\hsize=0.5\hsize}X
          }
\toprule
\textbf{Category} & \textbf{OpenSSL} &\textbf{PHP} \\
 \midrule
 Mention \weakness{} & 84 & 67 \\
 Mention explicit vulnerability  & 4 & 2 \\
 Others & 316 & 333 \\

 \bottomrule
\end{tabularx}
\newline
\label{table:weakness-prevalence}
\vspace{-6mm}
\end{table}


\subsection{PA2: Preliminary Evaluation of our Security Concern Identification Approach}

Since we cannot manually identify code review comments that contain \weakness{es} in the entire code review comment dataset (i.e., 135K comments; see Table \ref{table:dataset_size}), we opt to use a semi-automated approach to identify comments, as explained in Section~\ref{security_concern_identification_approach}. 
In particular, we measure the cosine similarity score of each code review comment and the descriptions of \weakness{} categories and we manually validate the comments with high cosine similarity scores until reaching the saturation point, i.e., 50 consecutive comments are identified as generic or irrelevant comments.  In this work, we explore two well-known vector representation techniques (i.e., TF-IDF and word embedding) when measuring cosine similarity.  We did not use the keyword search like prior works \citep{Bosu2014IdentifyingStudy, Paul2021AProject, Paul2021WhyProject} because their pre-defined keyword lists are limited and may not cover all coding weaknesses.
Hence, we set out this preliminary analysis to evaluate the effectiveness of our approach compared to the keyword search and examine which vector representation can produce the similarity scores that better distinguish the code review comments that contain \weakness{es} from the irrelevant code review comments.



\smallheading{Approach}
We conducted our preliminary evaluation based on the sampled dataset and our manual classification in PA1. We considered the comments that mentioned \weakness{} as \textit{\weakness{} comments} group, and the other comments as \textit{non-\weakness{} comments} group.
We pre-processed code review comments in the sampled dataset and the combined descriptions of \weakness{es} in all CWE-699 categories with the method described in Section~\ref{automated_concern_identification}. Then, we generated TF-IDF and word embedding vectors of the code review comments and the combined descriptions. Finally, we calculated the similarity score between the vectors.

To measure the effectiveness of our approach, we adopted the \textit{effort-aware} evaluation concept \citep{Kamei2013AAssurance, Verma2016OnRetrieval}.
\reviewersComment{R2.6}
\reviewersModification{
We measured top-$k$ precision, recall, and F1-score where $k$ is the number of comments with the highest similarity scores. 
While the value of $k$ approximates the effort required for our manual validation,
the top-$k$ precision shows the proportion of \textit{\weakness{} security comments} in the top-$k$ over the \textit{non-\weakness{} comments};
the top-$k$ recall shows the percentage of \textit{\weakness{} security comments} that can be identified at the top-$k$;
and the top-$k$ F1-score shows the single score that represent both top-$k$ precision and top-$k$ recall.
For the keyword search, we measured the precision, recall, F1-score and of the code review comments that were identified by a set of vulnerability keywords from previous secure code review studies~\citep{Bosu2014IdentifyingStudy, Paul2021AProject, Paul2021WhyProject}.
}

To evaluate the two vector representation techniques, we examine which technique produces similarity scores for \textit{\weakness{} comments} higher than the scores for \textit{non-\weakness{} comments}.
Thus, we used the one-sided Mann-Whitney-Wilcoxon test to examine the statistical difference in the similarity scores between the two groups of code review comments. We also used Cliff's $|\delta|$ effect size to estimate the magnitude of the difference in scores from each group.

\smallheading{Results}
\reviewersComment{R2.6}
\reviewersModification{
As shown in Table~\ref{table:approach_selection}, we found that our approach with word embedding vectors achieved the highest top-$k$ F1-score in OpenSSL and PHP for all $k$ $\in$ (20, 40, 60, 80, 100) with the top-$k$ F1-score of 0.16 - 0.58, while our approach with TF-IDF achieved the top-$k$ F1-score of 0.14 - 0.47.
Table~\ref{table:approach_selection} also shows that our approach achieves higher F1-score than the keyword search.
The keyword search retrieved 16 and 13 comments that contain one of the vulnerability keywords, which achieves an F1-score of 0.28 for OpenSSL and 0.25 for PHP. }
Moreover, we observe that the keyword search did not identify some types of \weakness{es} that can introduce vulnerability such as Pointer Issues (CWE-465). 
For example, the keyword approach could not identify a comment \textit{``The object can't be referenced after free\_obj, only dtor\_obj``}~\footnote{\url{https://github.com/php/php-src/pull/7079\#discussion_r660654029}} 
which is related to the `NULL Pointer Dereference` weakness (CWE-476).
This result shows that our approach using cosine similarity can identify more \weakness{} comments than the keyword search.



\begin{table*}[t]
\tiny
\centering\reviewersComment{R2.6}
\leftalignedcaption{Top-$k$ precision, Top-$k$ recall, and Top-$k$ F1-score of our approach, using TF-IDF and Word Embedding, and Precision, Recall, and F1-score of the keyword search approach. 
}
\begin{tabularx}{\linewidth}{
            >{\hsize=0.6\hsize}X
            >{\hsize=0.3\hsize}X
            >{\hsize=0.3\hsize}X
            >{\hsize=0.3\hsize}X
            >{\hsize=0.3\hsize}X
            >{\hsize=0.3\hsize}X
            >{\hsize=0.3\hsize}X
            >{\hsize=0.3\hsize}X
            >{\hsize=0.3\hsize}X
            >{\hsize=0.3\hsize}X 
            >{\hsize=0.3\hsize}X
            >{\hsize=0.3\hsize}X
            >{\hsize=0.3\hsize}X
            >{\hsize=0.3\hsize}X
            >{\hsize=0.3\hsize}X
            >{\hsize=0.3\hsize}X
            >{\hsize=0.3\hsize}X
            >{\hsize=0.3\hsize}X
            >{\hsize=0.3\hsize}X
          }
\toprule
\multirow{3}{*}{\textbf{Top-K}} &  \multicolumn{9}{c}{\textbf{OpenSSL}} & \multicolumn{9}{c}{\textbf{PHP}}\\
 \cline{2-19}
  & \multicolumn{3}{c}{\textbf{Keyword}} & 
  \multicolumn{3}{c}{\textbf{TF-IDF}} & 
  \multicolumn{3}{c}{\textbf{Word E.}} & 
  \multicolumn{3}{c}{\textbf{Keyword}} &
  \multicolumn{3}{c}{\textbf{ TF-IDF}} &  
  \multicolumn{3}{c}{\textbf{Word E.}}\\
  
  & \textbf{P} & \textbf{R} & \textbf{F1} & \textbf{P} & \textbf{R} & \textbf{F1} & \textbf{P} & \textbf{R} & \textbf{F1} & \textbf{P} & \textbf{R} & \textbf{F1} & \textbf{P} & \textbf{R} & \textbf{F1} & \textbf{P} & \textbf{R} & \textbf{F1} \\
  \hline
20 &  \multirow{5}{*}{ 0.53$^\dag$ } & \multirow{5}{*}{ 0.19$^\dag$ } & \multirow{5}{*}{ 0.28$^\dag$ }  & 0.60 & 0.14 & 0.23  & \textbf{0.80} & \textbf{0.19} & \textbf{0.31}  & \multirow{5}{*}{ 0.34$^\dag$ } & \multirow{5}{*}{ 0.20$^\dag$ } & \multirow{5}{*}{ 0.25$^\dag$ }   & 0.30 & 0.09 & 0.14  & \textbf{0.35} & \textbf{0.11} & \textbf{0.16}  \\

40 & & & & 0.50 & 0.24 & 0.32 & \textbf{0.70} & \textbf{0.33} & \textbf{0.45} & & & & 0.22 & 0.14 & 0.17 & \textbf{0.35} & \textbf{0.21} & \textbf{0.26} \\
60 & & & & 0.48 & 0.34 & 0.40 & \textbf{0.60} & \textbf{0.43} & \textbf{0.50} & & & & 0.23 & 0.21 & 0.22 & \textbf{0.42} & \textbf{0.38} & \textbf{0.40} \\
80 & & & & 0.45 & 0.43 & 0.44 & \textbf{0.56} & \textbf{0.54} & \textbf{0.55} & & & & 0.28 & 0.33 & 0.30 & \textbf{0.38} & \textbf{0.45} & \textbf{0.41} \\
100 & & & & 0.43 & 0.51 & 0.47 & \textbf{0.53} & \textbf{0.63} & \textbf{0.58} & & & & 0.29 & 0.47 & 0.35 & \textbf{0.36} & \textbf{0.54} & \textbf{0.43} \\

 \bottomrule 
  \multicolumn{19}{l}{{\tiny \dag 16 and 13 comments were retrieved from the keyword search for OpenSSL and PHP, respectively.}}
  
\end{tabularx}
\newline
\label{table:approach_selection}

\end{table*}

For the performance of similarity score calculation, Table~\ref{table:vector-evaluation} shows the results of the one-sided Mann–Whitney–Wilcoxon test and Cliff's $|\delta|$ effect size between the similarity scores of the \weakness{} comments and the non-\weakness{} comments.
We found that similarity scores of \weakness{} security comments are significantly higher than non-\weakness{} security comments ($p$-value $< 0.05$) when using TF-IDF and word embedding vectors.  
In addition, we found that the difference in the similarity scores from the word embedding vectors has a large effect size ($|\delta| \geq 0.474$ \citep{Romano2006AppropriateSurveys}) for both OpenSSL and PHP, while the difference in the similarity scores from TF-IDF vector has a large effect size for OpenSSL and a medium effect size for PHP. 
This suggests that the similarity scores based on the word embedding vectors can better differentiate \weakness{} comments from their counterparts than the similarity scores based on the TF-IDF vectors.
\reviewersComment{R2.6}
\reviewersModification{
This finding is consistent with the top-$k$ precision, recall and F1-scores shown in Table~\ref{table:approach_selection}, i.e., at the same $k$ value, using word embedding vectors achieves a higher score than using TF-IDF vectors. 
}

\begin{table}[hb]
  \centering
    \small
  \leftalignedcaption{Results of one-sided Mann–Whitney–Wilcoxon tests and Cliff's delta between the similarity scores of the \weakness{} and non-\weakness{} comments in the studied projects (n=400).}
  \begin{tabularx}{\linewidth}{
            >{\hsize=1\hsize}X
            >{\hsize=1\hsize}X
            >{\hsize=1\hsize}X
            >{\hsize=1\hsize}X
            >{\hsize=1\hsize}X
          }
    \toprule
    
    \multirow{2}{*}{\textbf{Project}} & \multicolumn{2}{c}{\textbf{TF-IDF}} & \multicolumn{2}{c}{\textbf{Word Embedding}} \\
    \cmidrule(lr){2-3} \cmidrule(lr){4-5}
    & MWW $p$-value &  Cliff's $|\delta|$ & MWW $p$-value  & Cliff's $|\delta|$   \\
    \midrule
    OpenSSL &  $1.41\mathrm{e}{-15}$ & 0.53895 (L) & $8.5\mathrm{e}{-25}$ & 0.67631 (L) \\ 
    PHP & $1.24\mathrm{e}{-9}$ & 0.45223 (M) &  $1.57\mathrm{e}{-13}$ & 0.54455 (L) \\
    
    \bottomrule
  \end{tabularx}
  \label{table:vector-evaluation}
\end{table}

Our preliminary evaluation shows that our approach with the word embedding technique 1) achieves a higher recall than the TF-IDF technique and the keyword search and 2) can better distinguish the \weakness{} comments. Therefore, in this study, we used the word embedding technique to calculate the similarity scores to help us manually identify the \weakness{} comments in the remaining dataset.

\section{Case Study Results}
\label{results}

\reviewersComment{R2.6}
\reviewersModification{
We report the empirical results based on the code review comments identified by the semi-automated approach; and answer the three research questions in this section, followed by a summary of our findings. 
}

\smallheading{RQ1: \rqOne} 
Table \ref{table:annotation_statistics} shows the number of identified code review comments and aggregated \securityConcern{s}.
From the 135K code review comments in the dataset, we manually read 3,570 OpenSSL and 2,576 PHP comments with the highest cosine similarity scores until reaching the saturation point (i.e., 50 consecutive irrelevant comments).
As described in Section~\ref{concern_manual_validation}, in the first iteration we removed irrelevant comments (e.g., related to bookkeeping and code styling), resulting in 232 and 148 comments. Subsequently, the first and the third author independently determined whether the comments raised legitimate security concerns and could be classified into one of the \weakness{} categories, resulting in 202 and 128 comments. To simplify the results, we aggregated comments within the same pull request that were classified into the identical \weakness{} category into singular \securityConcern{}.
In total, we identified 188 \securityConcern{s} from 202 comments in 164 pull requests in OpenSSL and 123 \securityConcern{s} from 128 comments in 100 pull requests in PHP. 
Note that one pull request can have multiple concerns with different \weakness{} categories. 
\reversemarginpar\reviewersComment{R3.6}
\reviewersModification{
The manual annotation process by the first and the third author achieved the inter-rater agreement~\citep{Cohen1960AScales} $\kappa= 0.70$ and $\kappa=0.84$ for OpenSSL and PHP, which can be interpreted~\citep{McHugh2012InterraterStatistic} as \textit{substantial} ($0.61 \geq |\kappa| \geq 0.81$) and \textit{almost perfect} ($ |\kappa| > 0.81$), respectively. }

\begin{table}[t]
\centering
\small
\leftalignedcaption{Numbers of code review comments that mentioned coding weaknesses and security concerns from our manual analysis.}
\begin{tabularx}{\linewidth}{
            >{\hsize=2.2\hsize}X
            >{\hsize=0.4\hsize}X
            >{\hsize=0.4\hsize}X
          }

\toprule
 \textbf{} & \textbf{OpenSSL} &\textbf{PHP} \\ [0.5ex] 
 \midrule
    Total comments read & 3,570 & 2,576 \\
\hline
    \textbf{Round 1 - pre-screening}: & & \\
    Preliminarily identified code review comments & 232 & 148 \\    
\hline
    \textbf{Round 2 - final validation}: &  &  \\
    Comments with legitimate \securityConcern{s} & 202 & 128 \\
    \hline
    \textbf{Identified security concerns} & 188 & 123 \\
 \bottomrule
\end{tabularx}
\newline
\label{table:annotation_statistics}
\end{table}

\begin{table}[t]
\centering
\small
\leftalignedcaption{Number of identified security concerns in CWE-699 categories.}
\begin{tabularx}{\linewidth}{
            >{\hsize=2.0\hsize}X
            >{\hsize=0.10\hsize}X
            >{\hsize=0.28\hsize}X |
            >{\hsize=0.10\hsize}X 
            >{\hsize=0.28\hsize}X
          }
\toprule
  \multirow{2}{*}{\textbf{Category (CWE-\#)}} & \multicolumn{2}{c|}{\textbf{OpenSSL}} & \multicolumn{2}{c}{\textbf{PHP}} \\
        & \# & \% & \# & \% \\
  \hline

    Authentication Errors (1211) \ddag&\textbf{21}&\textbf{11.17\%}&\textbf{9}&\textbf{7.32\%}\\
    API / Function Errors (1228) \ddag&\textbf{13}&\textbf{6.91\%}&\textbf{6}&\textbf{4.88\%}\\
    Random Number Issues (1213) \ddag&\textbf{13}&\textbf{6.91\%}&\textbf{5}&\textbf{4.07\%}\\
    Behavioral Problems (438 ) \ddag&\textbf{12}&\textbf{6.38\%}&\textbf{5}&\textbf{4.07\%}\\
    Resource Locking Problems (411)&\textbf{15}&\textbf{7.98\%}&3&2.44\%\\
    Cryptographic Issues (310) \ddag&\textbf{7}&\textbf{3.72\%}&\textbf{8}&\textbf{6.50\%}\\
    Audit / Logging Errors (1210)&5&2.66\%&\textbf{8}&\textbf{6.50\%}\\
    Initialization and Cleanup Errors (452)&\textbf{15}&\textbf{7.98\%}&1&0.81\%\\
    Concurrency Issues (557)&\textbf{9}&\textbf{4.79\%}&4&3.25\%\\
    Privilege Issues (265) \ddag&\textbf{7}&\textbf{3.72\%}&\textbf{5}&\textbf{4.07\%}\\
    Pointer Issues (465)&2&1.06\%&\textbf{8}&\textbf{6.50\%}\\
    Information Management Errors (199)&5&2.66\%&\textbf{6}&\textbf{4.88\%}\\
    Memory Buffer Errors (1218)&\textbf{7}&\textbf{3.72\%}&4&3.25\%\\
    Type Errors (136) &\multicolumn{2}{c|}{-}&\textbf{8}&\textbf{6.50\%}\\
    Resource Management Errors (399) &\multicolumn{2}{c|}{-}&\textbf{7}&\textbf{5.69\%}\\
    Business Logic Errors (840)&6&3.19\%&3&2.44\%\\
    Data Processing Errors (19)&6&3.19\%&2&1.63\%\\
    Communication Channel Errors (417) &\textbf{9}&\textbf{4.79\%}&\multicolumn{2}{c}{-}\\
    File Handling Issues (1219)&2&1.06\%&4&3.25\%\\
    Data Neutralization Issues (137) &\multicolumn{2}{c|}{-}&\textbf{5}&\textbf{4.07\%}\\
    Documentation Issues (1225) &\multicolumn{2}{c|}{-}&\textbf{5}&\textbf{4.07\%}\\
    String Errors (133)&1&0.53\%&4&3.25\%\\
    Handler Errors (429)&4&2.13\%&2&1.63\%\\
    Expression Issues (569) &\textbf{7}&\textbf{3.72\%}&\multicolumn{2}{c}{-}\\
    User Session Errors (1217)&2&1.06\%&3&2.44\%\\
    Data Integrity Issues (1214)&3&1.60\%&2&1.63\%\\
    Signal Errors (387)&2&1.06\%&2&1.63\%\\
    State Issues (371)&2&1.06\%&2&1.63\%\\
    Error Conditions, Return Values, ... (389) &4&2.13\%&\multicolumn{2}{c}{-}\\
    Permission Issues (275)&2&1.06\%&1&0.81\%\\
    Credentials Management Errors (255) &3&1.60\%&\multicolumn{2}{c}{-}\\
    Bad Coding Practices (1006) &2&1.06\%&\multicolumn{2}{c}{-}\\
    Data Validation Issues (1215)&\multicolumn{2}{c|}{-}&1&0.81\%\\
    Complexity Issues (1226) &1&0.53\%&\multicolumn{2}{c}{-}\\
    Key Management Errors (320) &1&0.53\%&\multicolumn{2}{c}{-}\\
    Authorization Errors (1212)&\multicolumn{2}{c|}{-}&\multicolumn{2}{c}{-}\\
    Encapsulation Issues (1227)&\multicolumn{2}{c|}{-}&\multicolumn{2}{c}{-}\\
    Lockout Mechanism Errors (1216)&\multicolumn{2}{c|}{-}&\multicolumn{2}{c}{-}\\
    Numeric Errors (189)&\multicolumn{2}{c|}{-}&\multicolumn{2}{c}{-}\\
    User Interface Security Issues (355)&\multicolumn{2}{c|}{-}&\multicolumn{2}{c}{-}\\

 \bottomrule
 \multicolumn{5}{l}{\tiny The bold text indicates the top 10 concerns in each project.} \\
 \multicolumn{5}{l}{\tiny ${\ddag}$ indicates the concerns that were frequently raised in both OpenSSL and PHP.}
\end{tabularx}
\newline
\label{table:rq1_top_concerns}
\end{table}

Table \ref{table:rq1_top_concerns} shows the number of identified \securityConcern{s} across the 40 \weakness{} categories of CWE-699.
The numbers in parentheses indicate the CWE category number of the \weakness.  
We found that in OpenSSL and PHP, identified \securityConcern{s} were related to 35 out of 40 \weakness{} categories of CWE-699, suggesting that diverse types of \weakness{es} can be discovered during the code review process.
The bold text in Table \ref{table:rq1_top_concerns} highlights the top ten \weakness{es} that were frequently raised in each project and the ${\ddag}$ symbol indicates the concerns that were frequently raised in both OpenSSL and PHP.
We found that six \weakness{es}, i.e., Authentication Errors (CWE-1211), API / Function Errors (CWE-1228), Privilege Issues (CWE-265), Behavioral Problems (CWE-438), Cryptographic Issues (CWE-310) and Random Number Issues (CWE-1213), were among the top ten concerns in both OpenSSL and PHP.
Additionally, we observe that several \weakness{es}  were frequently raised in a particular project.
This may suggest that while reviewers in OpenSSL and PHP share a set of common concerns, they can have a specific focus on particular security aspects as well.

Below, we present common \securityConcern{s} across both projects and project-specific \securityConcern{s}.

\begin{figure}[h]
    \centering
    \resizebox{\linewidth}{!}{\includegraphics{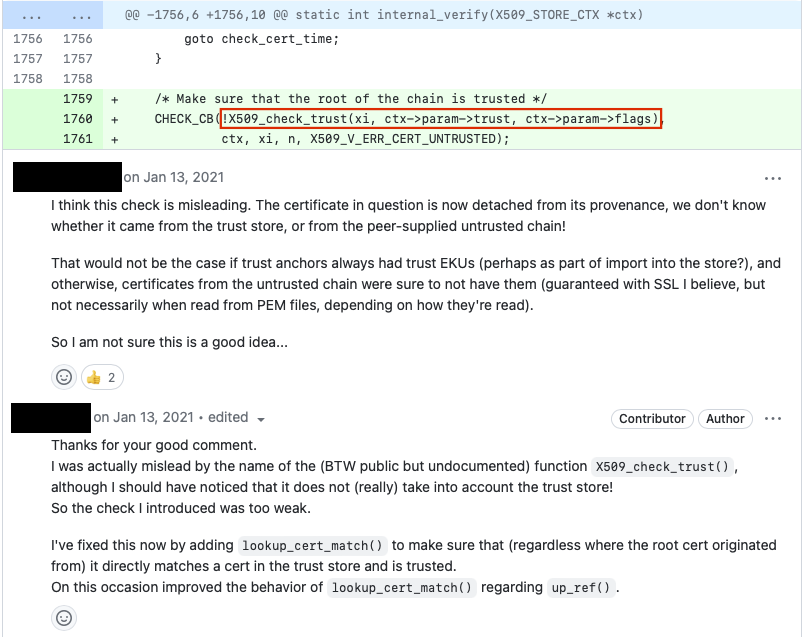}}
    \leftalignedcaption{A \securityConcern{} in category Authentication Errors (CWE-1211) in OpenSSL. }
    \label{fig:result-code-snippet-review-example-trusted-certificate}
\end{figure}

\inlineheading{Common \securityConcern{s} in OpenSSL and PHP:} 
The first two common \securityConcern{s} are related to users and rights, i.e., Authentication  Errors (CWE-1211) and Privilege Issues (CWE-265) \weakness{es}.
Authentication Errors (CWE-1211) are related to the failure to properly verify the identification of the rightful actors who can gain access to the system. 
For example, as shown in Figure~\ref{fig:result-code-snippet-review-example-trusted-certificate}, we observed that a reviewer noticed that the program does not verify whether the certificate is trusted or not: "\textit{[...]The certificate in question is now detached from its provenance, we don't know whether it came from the trust store, or from the peer-supplied untrusted chain![...]}".\footnote{\url{https://github.com/openssl/openssl/pull/13770\#discussion_r555847704}}
Privilege Issues (CWE-265) are related to the improper management of critical privileges assigned to users or objects.
For example, a reviewer mentioned that the developer did not use the correct approach to verify that the user has sufficient privileges to execute a script.\footnote{\url{https://github.com/php/php-src/pull/516\#issuecomment-27483079}}

\begin{figure}[h]
    \centering
    \resizebox{\linewidth}{!}{\includegraphics{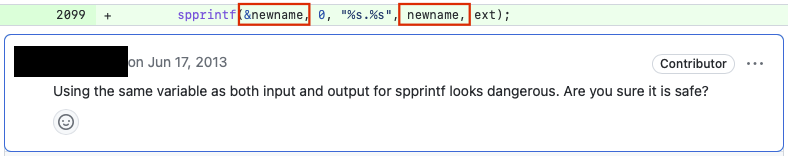}}
    \leftalignedcaption{A \securityConcern{} in category API/Function Errors (CWE-1228) in PHP.}
    \label{fig:result-code-snippet-review-example-ssprintf}
\end{figure}

Another two common \securityConcern{s} are related to \weakness{es} about the functionality of the system, i.e.,  API/Function Errors (CWE-1228) and Behavioral Problems (CWE-438).
API/Function Errors (CWE-1228) covers the use of dangerous functions or the exposing of the functions that allow unwanted actors to execute restricted actions. 
For example, as shown in Figure~\ref{fig:result-code-snippet-review-example-ssprintf}, we observed that a reviewer commented that assigning the result of the format string function to the same input variable can be potentially harmful: "\textit{[...]Using the same variable as both input and output for spprintf looks dangerous. Are you sure it is safe?}".\footnote{\url{https://github.com/php/php-src/pull/297/commits/062faf81e4a9e0f412c98a8309d42fb92e2a5737\#r4716852}}
Behavioral Problems (CWE-438) refer to code that may cause unexpected behavior in the software system. 
For example, a reviewer noticed that the code can look for the required files in incorrect directories if the program is compiled in different environments.\footnote{\url{https://github.com/php/php-src/pull/6146\#issuecomment-743779539}}

\begin{figure}[h]
    \centering
    \resizebox{\linewidth}{!}{\includegraphics{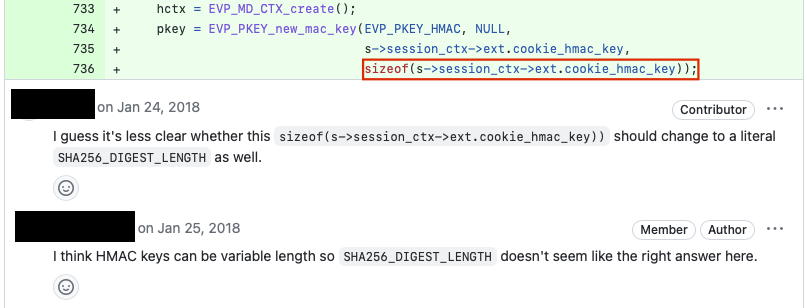}}
    \leftalignedcaption{A \securityConcern{} in category Cryptographic Issues (CWE-310). The developer had clarified the concern that was raised by a reviewer.}
    \label{fig:result-code-snippet-review-example-hmac-key-length}
\end{figure}

Concerns related to the cryptographic process, i.e., Cryptographic Issues (CWE-310) and Random Number Issues (CWE-1213), were also common in both OpenSSL and PHP.
Cryptographic Issues (CWE-310) covers the proper use of encryption algorithms and cryptographic keys to ensure system and data security. 
For instance, as shown in Figure~\ref{fig:result-code-snippet-review-example-hmac-key-length}, a developer responded to a reviewer's suggestion that the lengths of the cryptographic keys can be dynamic and cannot be restricted to a fixed value by saying "\textit{[...]HMAC keys can be variable length so SHA256\_DIGEST\_LENGTH doesn't seem like the right answer here}".\footnote{\url{https://github.com/openssl/openssl/pull/4435\#discussion_r163626967}}
Random Number Issues (CWE-1213) account for the process of obtaining sufficient randomness, which is essential for robust data encryption.
For example, a reviewer suggested that the library should have an inlet for external entropy to increase randomness in the random number generator.\footnote{\url{https://github.com/openssl/openssl/pull/7399\#issuecomment-432535460}}

Including the six common \weakness{es}, there are 21 types of \weakness{es} that were raised in both projects.
In particular, \securityConcern{s} related to \weakness{es} in category Audit/Logging Errors (CWE-1210), Information Management Errors (CWE-199), Concurrency Issues (CWE-557), Memory Buffer Errors (CWE-1218), Business Logic Errors (CWE-840), and Resource Locking Problems (CWE-411) are among the top 20 categories in both projects. \securityConcernC{s} in these categories may also be considered common concerns to some extent. 

\reviewersModification{
\reviewersComment{R2.11}
The previous code review works~\citep{Alfadel2023EmpiricalPackages, Paul2021WhyProject, DiBiase2016AChromium, Bosu2014IdentifyingStudy, Edmundson2013AnReview} reported that reviewers can identify \securityIssue{s} in various degrees based on the different application domains and the programming languages.
However, the studied \securityIssue{s} are frequently bounded by well-known vulnerabilities that are associated with security consequences such as SQL Injection, XSS, or Denial of service.
Our results further reveal that reviewers can commonly discuss more extensive \weakness{es} that can introduce those vulnerabilities from the development perspective.
For example, the discussion regarding API / Function Errors (CWE-1228), Behavioral Problems (CWE-438), Cryptographic Issues (CWE-310), and Random Number Issues (CWE-1213) have not been previously reported.
}

\begin{tcolorbox}[colback=gray!5!white,colframe=gray!100!black]
  \textbf{Summary}: While various types of \weakness{es} (35 out of 40 categories) can be raised during code reviews, \textbf{Authentication Errors (CWE-1211)}, \textbf{Privilege Issues (CWE-265)}, \textbf{API / Function Errors (CWE-1228)}, \textbf{Behavioral Problems (CWE-438)}, \textbf{Cryptographic Issues (CWE-310)}, and \textbf{Random Number Issues (CWE-1213)} were the most frequently raised in OpenSSL and PHP.
\end{tcolorbox}

\inlineheading{Project-specific \securityConcern{s}:} 
In addition to common \securityConcern{s}, understanding project-specific concerns would allow us to gain better insight into the secure code review practices in each project. 
We observed that in OpenSSL, a library that provides encryption functionalities to its dependent systems, reviewers seem to focus on preventing direct security threats that are related to encryption, e.g., Key Management Errors (CWE-255) and Communication Channel Errors (CWE-417).
For example, a reviewer discussed the causes of \textit{timing-attack}, which can reveal the type of cryptographic key used in secure communication with the attacker.\footnote{\url{https://github.com/openssl/openssl/pull/16944\#issuecomment-957022300}}

On the other hand, in PHP, a programming language for web applications, reviewers rather focus on security related to data controlling, e.g., Data Validation Issues (CWE-1215) and the versatility of language, e.g., Pointer Issues (CWE-465) and Type Errors (CWE-136).
Also, it seems that PHP reviewers are concerned with Documentation Issues (CWE-1225), which are rarely recognized in a security context~\citep{Alfadel2023EmpiricalPackages}.
For example, a developer explained to a reviewer that a function should not declare to accept any type of parameters if it intends to raise \texttt{TypeError} when the user inputs the parameters of incorrect types, e.g., to avoid Denial of Service \vulnerability{y}.\footnote{\url{https://github.com/php/php-src/pull/5847\#discussion_r454239763}} In another case, a reviewer noticed that a function does not implement a randomization algorithm that it claims to use in the document.\footnote{\url{https://github.com/php/php-src/pull/1681\#issuecomment-187120620}} 
These types of \securityConcern{s} highlight the importance of input management and documentation in PHP.

Lastly, for the \weakness{} types that were rarely raised, it may be because these issues are irrelevant to the application domains of the systems. 
We did not observe any concerns related to Lockout Mechanism Errors (CWE-1216), as it can cause an overly restrictive authentication policy, which is not applicable in both projects.
Similarly, no concerns related to User Interface Security Issues (CWE-355) were found, as OpenSSL and PHP do not have an elaborate user interface. Therefore, it is less likely that reviewers would raise this type of concern.

\begin{tcolorbox}[colback=gray!5!white,colframe=gray!100!black]
  \textbf{Summary}: A software project can have particular \securityConcern{s} from \weakness{es}. In addition to common security concerns across the two studied projects, OpenSSL has additional concerns related to \textbf{direct security threats} which are associated with encryption and PHP has additional concerns related to \textbf{data controlling and the versatility of language} such as data type and pointer.
\end{tcolorbox}

\smallheading{RQ2: \rqTwo} 
\label{rq2_result}
Based on the mapping of \knownV{} \vulnerability{ies} to related \weakness{es}, as explained in Section~\ref{mapping_past_vulnerability}, we find that the \knownV{} \vulnerability{ies} of OpenSSL and PHP during the studied period are related to 16 \weakness{} categories.
\reviewersComment{R3.9}
\reviewersModification{
We answer this question by comparing the percentages of the \knownV{} \vulnerability{ies} and the raised \securityConcern{s} that we found in RQ1 (Table~\ref{table:rq1_top_concerns}).}


\begin{figure}[h]
    \centering
    \resizebox{\linewidth}{!}{\includegraphics{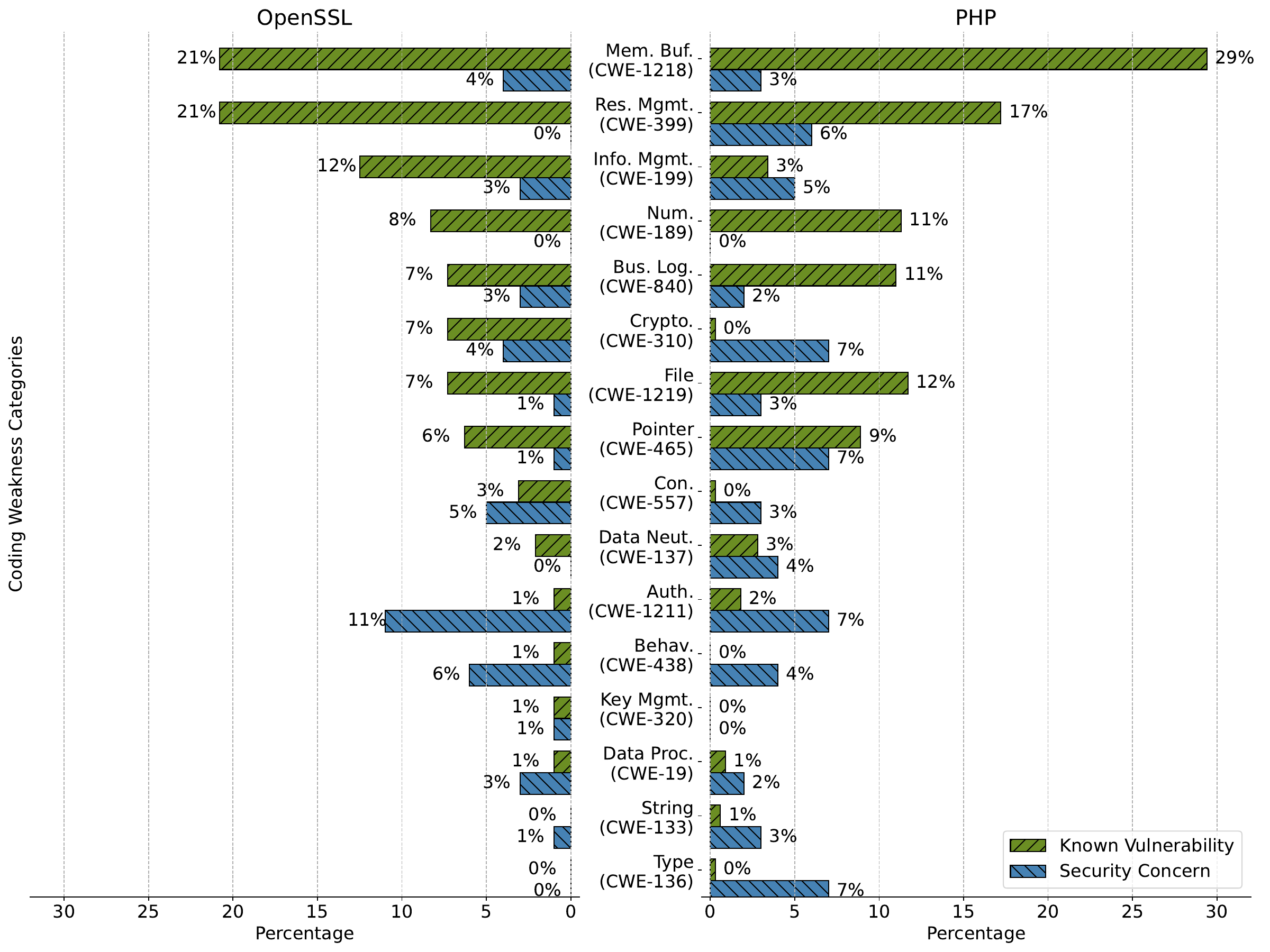}}
    \leftalignedcaption{The distribution of \knownV{} \vulnerability{ies} and the \securityConcern{s} across the \weakness{} categories (the CWE-699 taxonomy). Note that the other categories that did not occur in \knownV{} \vulnerability{ies} are omitted. 
    }
    \label{fig:req-past-vulnerability-alignment}
\end{figure}

Figure~\ref{fig:req-past-vulnerability-alignment} shows that nine \weakness{} categories in OpenSSL and six \weakness{} categories in PHP have a high proportion of \knownV{} \vulnerability{ies} in the past, but are less frequently discussed in code reviews. 
For instance, the top two \weakness{} categories that have the highest proportion of \knownV{} \vulnerability{ies} are Memory Buffer Errors (CWE-1218; 21\% in OpenSSL and 29\% in PHP) and Resource Management Errors (CWE-399; 21\% in OpenSSL and 17\% in PHP).
However, these two \weakness{} categories have a relatively low proportion of security concerns raised in the code reviews (4\% - 9\%).
Similarly, 6\% - 12\% of the \knownV{} \vulnerability{ies} are related to Business Logic Errors (CWE-840), File Handling Errors (CWE-1219), and Pointer Issues (CWE-465) which were rarely discussed in the code review (only 1\% - 7\% of the \securityConcern{s}).

Moreover, we observe that OpenSSL has three \weakness{es} that are less-frequently discussed in code reviews i.e., Information Management Errors (CWE-199) (17\% of \knownV{} \vulnerability{ies}; 3\% of \securityConcern{s}), Cryptographic Issues (CWE-310) (7\% of \knownV{} \vulnerability{ies}; 4\% of \securityConcern{s}), and Data Neutralization Issues (CWE-137) (2\% of \knownV{} \vulnerability{ies}; 0\% of \securityConcern{s}). 
In particular, the lower number of \securityConcern{s} about Data Neutralization Issues align with the observation of~\cite{Braz2021Why-} that developers may not be aware of the consequences of improper input validation, as well as the case of \textit{Heartbleed} as shown in Figure~\ref{fig:motivating-example-heartbleed-fix}. 

On the other hand, \weakness{es} in six categories in both OpenSSL and PHP were more frequently discussed than the \knownV{} \vulnerability{ies}. \weaknessC{es} related to Authentication Errors (CWE-1211), String Errors (CWE-133), Type Errors (CWE-136), Concurrency Issues (CWE-557), Data Processing Errors (CWE-19), and Behavioral Problems (CWE-438) which occurred in 4\% of \knownV{} \vulnerability{ies} in OpenSSL and PHP were discussed by 22\%-23\% of \securityConcern{s} in both projects. 


Despite the low frequency of the \securityConcern{s} compared to the \knownV{} \vulnerability{ies}, all of the \weakness{} categories of the \knownV{} \vulnerability{ies}, except for Numeric Errors (CWE-189)  were discussed in the code review as shown in Figure~\ref{fig:req-past-vulnerability-alignment}.
This finding suggests that reviewers may be able to identify these kinds of \weakness{}, but require more attention.


\begin{tcolorbox}[colback=gray!5!white,colframe=gray!100!black]
  \textbf{Summary}: \weaknessC{es} related to \textbf{memory},  \textbf{resource management}, \textbf{numeric errors}, \textbf{business logic}, \textbf{file handling}, and \textbf{pointer} are less frequently discussed compared to the \knownV{} \vulnerability{ies} in both projects. 
  Nevertheless, almost every type of \weakness{es} related to \knownV{} \vulnerability{ies} can be identified and discussed in code reviews.
\end{tcolorbox}

\smallheading{RQ3: \rqThree} 
Through our qualitative analysis of code review activities in pull requests with \securityConcern{s} described in Section \ref{handling_process_identification}, we identified eight scenarios in which \securityConcern{s} were handled, and we grouped them into four main themes.
Table \ref{table:response_scenario_distribution} shows the number of concerns in each handling scenario.\footnote{Note that we excluded two \securityConcern{s} in PHP as we could not find the discussions that are specific to these concerns in the pull requests.}
Below, we describe each handling scenario.



\begin{table}[t]
\centering
\small
\leftalignedcaption{Identified security concerns' scenarios and their distribution. The complete data set, including scenario distribution by category, can be found in the data release package. }
\begin{tabularx}{\linewidth}{
            >{\hsize=0.1\hsize}X
            >{\hsize=2.3\hsize}X
            >{\hsize=0.8\hsize}X
            >{\hsize=0.8\hsize}X
          }
\toprule

   \multicolumn{2}{l}{\multirow{2}{*}{\textbf{Security concerns' response scenario}}} & \textbf{OpenSSL} & \textbf{PHP}\\
  & & \# (\%) & \# (\%) \\
  \hline
\multicolumn{2}{l}{\textbf{C1: Fix attempted}}  & \textbf{78 (41\%)} & \textbf{48 (39\%)} \\
    & C1.1: Successfully fixed & 70 (37\%)  & 45 (37\%)\\
    & C1.2: Unsuccessfully fixed & 8 (4\%)  & 3 (2\%)\\
 \hline
 \multicolumn{2}{l}{\textbf{C2: Acknowledged }} & \textbf{56 (30\%)} & \textbf{44 (36\%)}\\
    & C2.1: Fix elsewhere & 20 (10\%)  & 22 (18\%)\\
    & C2.2: Unresolved & 36 (20\%)  & 22 (18\%) \\
 \hline
 \multicolumn{2}{l}{\textbf{C3: Dismissed}} & \textbf{49 (26\%)} &  \textbf{18 (15\%)}\\
    & C3.1: False concern & 25 (13\%)  & 10 (8\%) \\
    & C3.2: Design choice & 24 (13\%)  & 8 (7\%) \\
 \hline
 \multicolumn{2}{l}{\textbf{C4: Unresponded }} & \textbf{5 (3\%)} &\textbf{11 (9\%)}  \\
    
    & C4.1: Out of context & --  & 2 (2\%) \\
    & C4.2: Unknown & 5 (3\%)  & 9 (7\%)  \\
 \bottomrule
\end{tabularx}
\newline
\label{table:response_scenario_distribution}
\end{table}


\inlineheading{C1. Fix attempted (39\% in OpenSSL; 41\% in PHP):}
For the largest group of \securityConcern{s}, we observed that developers attempted to address the concern by modifying code (i.e., committing new changes to the pull request).
However, the attempt can be either successfully fixed (C1.1; 37\% for both projects) or unsuccessfully fixed (C1.2; 4\% for OpenSSL and 2\% for PHP).
The successfully fixed scenario (C1.1) refers to cases where reviewers accepted to merge the pull request after the code was modified.
For example, a reviewer requested that the developer adjust the memory handling process.\footnote{\url{https://github.com/openssl/openssl/pull/7611\#discussion_r238410834}} The developer further inquired and eventually modified the code and committed a new change to the same pull request.
The unsuccessfully fixed scenario (C1.2) refers to cases where reviewers declined or did not respond to the new changes that the developers made, which eventually led to pull request rejection.
For example, reviewers suggested an alternate approach for cleaning up the memory; the developer made a fix, but never received further feedback.\footnote{\url{https://github.com/openssl/openssl/pull/5495\#issuecomment-370178440}}

\inlineheading{C2. Acknowledged (30\% in OpenSSL; 36\% in PHP):} \label{C2: Acknowledge}
For a third of the \securityConcern{s} raised, we observed that \securityConcern{s} were acknowledged by the developer or other reviewers but were not fixed in the same pull request.
We observed that the concerns were not fixed in the same pull request because they will be fixed elsewhere (C2.1; 10\% for OpenSSL and 18\% for PHP) or due to an unresolved discussion (C2.2; 20\% for OpenSSL and 18\% for PHP).
In particular, for the fix-elsewhere scenario (C2.1), the reviewers and developers discussed the raised concern and agreed that the necessary fixes should be made in new pull requests.
We find that around half (55\%) of the \securityConcern{s} in this scenario were eventually merged in both projects ($\frac{11}{20}$ for OpenSSL and $\frac{12}{22}$ for PHP).
For example, a reviewer noticed the use of \textit{stale pointer} and suggested a fix. The developer then replied, "\textit{[...] Ok. I'll prepare a pull request (but not right away) and request your review.}".\footnote{\url{https://github.com/openssl/openssl/pull/7455\#discussion_r227727671}}
\reviewersComment{R3.7}
\reviewersModification{
However, it is not possible to confirm whether all \securityConcern{s} in the C2.1 scenario were later fixed as promised.
}

For the unresolved discussion scenario (C2.2), the developers and reviewers cannot find an agreeable direction to address the concern. 
The discussions in this scenario tend to be more rambling and involve several sub-concerns, hindering reviewers from reaching an agreeable resolution.
This could be due to different understandings and perspectives between reviewers.
For example, reviewers and developers discussed the resolution while aiming to maintain compliance with security standards. However, due to the equivocal interpretation of the standards, the discussion cannot reach an agreeable resolution.\footnote{\url{https://github.com/openssl/openssl/pull/5402\#issuecomment-369513189}}
Another example is that a reviewer raised a concern about the certificate authentication process and requested a modification.\footnote{\url{https://github.com/openssl/openssl/pull/4848\#issuecomment-363287772}} 
The other reviewers, including the developer, agreed that the concern was valid but expressed multiple opinions on the solutions. 
The pull request with the concern was eventually merged without any changes.
\reviewersComment{R3.7}
\reviewersModification{
Indeed, we found that 16 pull requests in OpenSSL and 5 pull requests in PHP which contain 16 ($\frac{16}{36}=44\%$ for OpenSSL) and 7 ($\frac{7}{22} = 31\%$ for PHP) concerns in C2.2 were eventually merged without any evidence that the concerns were addressed.
}

\reviewersComment{R1}
\reviewersModification{
It should be noted that the reviewer's workload may affect the code review outcomes.
We found that a significant portion of reviewers (54\% in OpenSSL and 17\% in PHP) engaged in unresolved discussions (C2.2) are classified as high-workload reviewers i.e., reviewed over 100 pull requests in each respective project.
We hypothesize that workload, characterized by the volume of code reviews, as discussed in prior research~\citep{Ruangwan2019TheReview}, could influence the quality of code review process.
However, future work can be conducted to further investigate this phenomenon.
}


\inlineheading{C3. Dismissed (15\% in OpenSSL; 26\% in PHP):} 
In this scenario, the developer and reviewers discussed the \securityConcern{s} raised, and the \securityConcern{s} were dismissed. 
We observed that the discussions eventually concluded that the concern was a false concern (C3.1; 13\% for OpenSSL and 7\% for PHP) or acceptable by design choice (C3.2; 24\% for OpenSSL and 7\% for PHP). 
Specifically, the false concern scenario (C3.1) is related to cases in which developers or other reviewers offered an explanation to invalidate the \securityConcern{s}.
For example, a reviewer raised a concern about leaking sensitive data.\footnote{\url{https://github.com/php/php-src/pull/5340\#discussion_r402406437}}
Then, the developer replied to the comment to explain that the implementation is not leaking sensitive data "\textit{[...] \%s given part shouldn't be added for values (but only for types) since they might contain sensitive data}", which was agreed by the reviewer.
The design choice scenario (C3.2) refers to cases where \securityConcern{s} were dismissed by other factors such as performance trade-off, maintainability, or system design~\citep{Zanaty2018AnReview}.
For example, a developer responded that a change in the data-neutralizing process was a valid concern as raised by the reviewer; however, it did not affect the application logic.\footnote{\url{https://github.com/openssl/openssl/pull/12251\#discussion_r445535130}} The reviewer finally agreed and approved the pull request.
\reviewersComment{R3.7}
\reviewersModification{
We also observed that 20 pull requests in OpenSSL and 4 pull requests in PHP which contain 21 ($\frac{21}{24}=88\%$ for OpenSSL) and 4 ($\frac{4}{8}=78\%$ for PHP) concerns in scenario C3.2 were eventually merged.
}

\inlineheading{C4. Unresponded (3\% in OpenSSL; 9\% in PHP):} 
There were a few cases where \securityConcern{s} did not receive any responses nor activities logged in the pull request.
This was in part due to out-of-context (C4.1; 2\% for PHP) or unknown \& inactivity (C4.2; 3\% for OpenSSL and 7\% for PHP). 
The out-of-context scenario (C4.1) refers to cases where \securityConcern{s} drift away from the current discussion or the goal of the code changes.
For example, a reviewer raised a \securityConcern{} about insufficient check of input and instantly volunteered to create a new change request that fixes the problem, however, the developer and other reviewers did not respond to the concern.\footnote{\url{https://github.com/php/php-src/pull/8184\#issuecomment-1069803761}} 
The unknown \& inactivity scenario (C4.2) refers to cases where \securityConcern{s} were simply disregarded without a clear reason.
For example, a reviewer remarked suspicious use of pointer but the developer did not respond and the pull request was eventually rejected.\footnote{\url{https://github.com/php/php-src/pull/354/commits/c582b3d5cd986002c242adae54d1dfd7ee26ff11\#r6856182}}


\reviewersComment{R3.8}
\begin{tcolorbox}[colback=gray!5!white,colframe=gray!100!black]
\reviewersModification{\textbf{Summary}: For many of the \securityConcern{s} raised, the developers \textbf{attempted to fix them within the same pull request}. 
However, 30\%-36\% of the \securityConcern{s} raised were only \textbf{acknowledged and merged without immediate fixes} and 15\%-26\% of the concerns were \textbf{dismissed} or \textbf{not responded to}.}
\end{tcolorbox}

\section{Discussion}
\label{discussion}

In this section, we discuss the implications of our results and provide practical recommendations for practitioners and potential future work.

\smallheading{1) Various \weakness{es} that may lead to \securityIssue{s} can be raised during code reviews.}
Our first preliminary analysis (PA1) in Section~\ref{preliminary_analysis} shows that \weakness{es} were raised in the code review process 21 - 33.5 times more often than explicit \vulnerability{ies}. 
This finding supports our intuition that the reviewers tend to focus on issues in source code. 
Therefore, it is more natural for the reviewers to identify \weakness{es} than \securityIssue{s}. 
This implication aligns with the previous work~\citep{Goncalves2022DoLoad} that the cognitive load required for code reviews is lower if the reviewers already have the relevant knowledge.

Indeed, our RQ1 shows that the raised \securityConcern{s} in code reviews of OpenSSL and PHP cover nearly 90\% of the CWE-699 weakness types (i.e., 35 out of 40 categories, see Table~\ref{table:rq1_top_concerns}). This confirms our presumption that a variety of \weakness{es} can be raised by reviewers during the code review process. As shown in the motivating examples in Section~\ref{motivation}, such \weakness{es} can lead to \securityIssue{s}.  It can be implied that the \weakness{es} that may introduce \securityIssue{s} can potentially be identified during the code review process although the weaknesses did not yet explicitly expose the vulnerable outcomes~\citep{Braz2021Why-}. 
Our manual observations from RQ1 also show that the code changes may potentially be vulnerable if the author did not address the raised \securityConcern{s}. For instance, Figure~\ref{fig:result-code-snippet-review-example-trusted-certificate} shows that \vulnerability{ies} such as CVE-2008-4989\footnote{\url{https://nvd.nist.gov/vuln/detail/CVE-2008-4989}} and CVE-2012-5821\footnote{\url{https://nvd.nist.gov/vuln/detail/CVE-2012-5821}} could be introduced into the code if the Improper Certificate Validation \weakness{} (CWE-295) under the Authentication Errors category (CWE-1211) was not raised by a reviewer.

\reviewersComment{R2.11}
\reviewersModification{
\underline{Recommendation:} As we found that \weakness{es} can be identified in code reviews, our findings suggest that practitioners and/or other software projects could adopt the \weakness{es} taxonomy (i.e., CWE-699) to assist code reviews. 
A list of \weakness{es} should help the team increase the awareness of the potential problems that can lead to \securityIssue{s} without requiring deep security knowledge. 
A recent controlled experiment of \cite{Braz2022LessFormat} has shown that a code review checklist could help reviewers better find \securityIssue{s}. 
Hence, one of the possible ways to adopt the \weakness{es} taxonomy for code reviews is to incorporate it into a code review checklist. 
Future work should investigate the effectiveness and practicality of using \weakness{es} as a code review checklist for identifying and mitigating \securityIssue{s} during the code review process. 
Moreover, as \weakness{} are more frequently discussed than the \securityIssue{}, \weakness{} can also be an effective proxy for understanding secure code review practices.\\
}


\smallheading{2) \weaknessC{es} related to the \knownV{} \vulnerability{ies} of the systems are not frequently discussed in code reviews.}
Our RQ2 shows that some types of \weakness{es} were less frequently discussed compared to the \knownV{} \vulnerability{ies} (see Figure~\ref{fig:req-past-vulnerability-alignment}). 
In particular, we found that Memory Buffer Errors (CWE-1218) and Resource Management Errors (CWE-399) are the least frequently discussed \weakness{es} in OpenSSL and PHP (4\%-9\%), albeit the high percentages of \knownV{} \vulnerability{ies} (17\%-29\%).
Furthermore, our motivating examples in Section~\ref{motivation} highlighted that such \weakness{es} can lead to a serious \vulnerability{y}. 
For example, OpenSSL's \textit{Heartbleed} is a \knownV{} \vulnerability{y} related to weakness Out-of-bounds Read (CWE-125) which is a type of memory buffer error. 
These \weakness{es} were rarely discussed maybe because they are generic and easy to be overlooked. Hence, the reviewers may have failed to notice them. 
To mitigate this problem, the reviewers should be aware of these latent \weakness{es} in order to properly prioritize them in the code reviews.


In addition to the \knownV{} \vulnerability{ies}, our RQ1 indicates that the \securityConcern{s} in code reviews can vary from project to project. 
Particularly, OpenSSL reviewers were concerned about direct security threats (e.g., Authentication Errors (CWE-1211 and Random Number Issues (CWE-1213)), while PHP reviewers were more concerned about data controlling (e.g., Type Errors (CWE-136)). 
As OpenSSL is an encryption library for secure communication and PHP is a programming language, it can be implied that the application domain may correlate with the \weakness{es} that reviewers can raise.
This finding also supports our results that \weakness{es} such as User-interface Security Issues (CWE-355) and Encapsulation Issues (CWE-1227) were neither found in our results nor appear in the \knownV{} \vulnerability{ies} because they are less related to the application domains of the studied projects. 

\underline{Recommendation:} Our findings suggest that it is essential to identify the specific \weakness{es} that are significant, highly prone to introduce \securityIssue{s}, and relevant to the application domain of the projects.
Thus, rather than reviewing all types of \weakness{es}, a selected set of \weakness{es} can be prioritized for effective code reviews.
Prioritization of \weakness{es} during code reviews can be based on \knownV{} \vulnerability{ies} and the unique concerns of the projects that were raised in the past. 
Future work can investigate a systematic approach for identifying and prioritizing the types of important \weakness{es} for individual projects in this context.\\

\smallheading{3) Not all the raised \securityConcern{s} were addressed within the same code review process.} 
The \securityConcern{} handling scenarios identified in our RQ3 reveal a shortcoming in the code review process. Our results show that approximately a third of the \securityConcern{s} from \weakness{es} (30\%-36\%, see C2 in Table~\ref{table:response_scenario_distribution}) were acknowledged without fixes in the process.
We observed that developers promised to fix some of the acknowledged concerns in the new independent code changes (10\%-18\%), but some concerns were left without fixing due to disagreement about the proper solution (18\%-20\%).
Nevertheless, approximately half of the unresolved concerns (6\%-9\%) were eventually merged. 
This result implies a possible risk that \securityIssue{s} can slip through the code review process into the software product. 
The incomplete code reviews or unclean code changes that contain \securityConcern{s} related to \weakness{es} should be held from merging until all \securityConcern{s} are resolved. 
Otherwise, the remaining \weakness{es} in code changes can become \securityIssue{s} in the future. 
This implication is consistent with the findings of the prior work which reported that relentless and inconclusive discussion could impact the code review quality~\citep{Kononenko2015InvestigatingMatter}, and the incomplete code reviews and the unsuccessfully fixed can negatively affect the developer's contribution~\citep{Gerosa2021TheSource}. 

\underline{Recommendation:} Code reviews with \securityConcern{s} should be escalated if the final resolutions cannot be agreed upon before merging. Security experts or experienced developers should be included in such code reviews to investigate complex \securityConcern{s}.
In addition, the mechanisms to notify the reviewers of the incomplete code reviews or the insufficiently addressed \securityConcern{s} could reduce the risk that \securityIssue{s} will slip through the code review process into the software product.
Our suggestion aligns with \cite{Wessel2020EffectsProjects} who reported that the adoption of an automated mechanism such as code review bots can increase the number of merged pull requests, and, hence, reduce the number of abandoned code reviews.
\cite{Kudrjavets2022MiningAnalysis} also observed that the automated bots can remind the developers of the pending tasks in the code review process without inciting negative feelings.
Hence, future work should investigate an approach to identify incomplete code reviews or the insufficiently addressed \securityConcern{s} to help developers increase awareness.

\section{Threats to Validity}
\label{threats}

We discuss potential threats to the validity of our study.


\subsection{Internal Validity}
\reviewersComment{R1.3, R2.8}
\reviewersModification{
During the manual annotation to identify \securityConcern{s}, code review comments can be ambiguous or require more contextual information to understand.
In such cases, we decided to preserve the precision of the manual annotation by considering the ambiguous or unclear-context comments as irrelevant to \weakness.
However, as the annotation process was conducted categorically, it may be susceptible to the biases of the annotator.
To mitigate this, the comments were independently validated by the third author (Section~\ref{concern_manual_validation}).
Additionally, if the comments are relevant to multiple categories (i.e., receiving high similar scores in multiple categories), they were also annotated and validated multiple times.
During the validation of handling scenarios in RQ3 (see Section~\ref{handling_process_identification}), we encountered a few instances of disagreement. 
We attribute this discrepancy to the limitations inherent in code review data and a potential lack of expertise in the project.
}
We were aware that some weaknesses in the CWE-699 taxonomy are not considered harmful from a security perspective. Thus, we regularly consulted the extended description in CWE-699 to ensure that the \securityConcern{s} in question can lead to vulnerabilities.
We were also aware that several categories in CWE-699 may share similar weaknesses. For example, weaknesses in the Random Number Issues (CWE-1213) category are also listed in the Cryptographic Issues (CWE-310) category. Nevertheless, we only identified three \securityConcern{s} that shared both \weakness{es}.

\subsection{Construct Validity}
We used an automated text-based approach to facilitate our manual annotation process. The performance of the automated approach can be suboptimal due to the limited vocabulary in the documents. We tried to mitigate this concern by including CWE's alternate terms that developers might use.
It should also be noted that the selection of word-embedding techniques can impact the possibility of finding relevant code review comments. We carefully selected the word-embedding model pre-trained in the software engineering domain to reduce the potential issues.
In the manual annotation process, we read only comments that have high similarity scores (i.e., reading and doing manual analysis until reaching the saturation point). It is possible that some of the unread comments may also contain \weakness{es}.

For RQ2, we analyzed the alignment of \knownV{} \vulnerability{ies} and \securityConcern{s} by observing the distribution of related weaknesses. It is worth noting that CWE assignments for CVE are based on the security expert's judgment. Therefore, they can be subjective. Additionally, CVE records can be updated. Hence, our analysis is limited by the abstract observations at the time of data collection. 
For RQ3, we found two PHP pull requests with a long thread of discussions (100-300 comments). Although we were able to locate the identified comments, it is difficult to observe the handling scenarios, i.e., whether the issue was eventually addressed by developers or not. To avoid misinterpretation of the handling process based on these code review activities, we decided to drop these two pull requests from the results of our RQ3. We tried to minimize this problem by manually checking the final code change and the developer's reactions. However, there is no effective solution to completely mitigate this issue.
For transparency, we released the dataset used in this study in our supplementary materials. 

\reviewersComment{R2.7}
\reviewersModification{
Finally, the quality of the studied datasets can affect the validity of the results. Although the studied projects primarily conduct code reviews on GitHub, we cannot guarantee that our datasets include every code review in each project because some code reviews may not be documented.
}\\
    
\subsection{External Validity}
\reviewersComment{R2.10}
\reviewersModification{
While increasing the number of studied projects may strengthen the generalizability of the findings, expanding the studied subjects is not a trivial task.
This is because there are a limited number of projects that fit our selection criteria e.g., the size of projects (the small projects may not have sufficient security discussion~\citep{DiBiase2016AChromium}), the past vulnerabilities (for comparing the alignment of past vulnerabilities), the availability of code review data, and the mandatory code review policy. 
Furthermore, \citet{Nagappan2013DiversityResearch} also suggested that indiscriminately increasing the sample size in software engineering study may not necessarily improve the generalizability.
}

During the annotation process, we observed that both studied projects have several special traits due to a different application domain. 
The findings based on these two projects may include aspects that may not apply to other software projects. 
Thus, the analysis of the studied dataset does not allow us to draw conclusions for all open-source projects. Nevertheless, we carefully selected two distinct projects for this study that differ in nature and potential \securityIssue{s}. PHP is a general-purpose scripting language that may face a wide range of varying levels of security threats depending on its usage. OpenSSL is a library with a primary focus on security. Hence, we believe that \securityIssue{s} present in both of these projects are also relevant to other software projects within similar application domains. 
Further studies are required to confirm this hypothesis.

\reviewersComment{R2.9, R2.12}
\reviewersModification{
As our findings are based on the snapshot of code review datasets until June 2022, the recency of the data can be a concern.
To mitigate this issue, we analyzed the \weakness{es} in the newly collected code review datasets between June 2023 and February 2024 from both projects, which comprise 6,365 code review comments and 1,427 pull requests in total. 
We found no major difference in the prevalence of \weakness{} discussion between the two datasets. 
In particular, nine categories remain in the top 10 categories of OpenSSL, and six categories remain in the top 10 categories of PHP.~\footnote{The annotated code review datasets from the extra analysis are included in the supplementary material.}
However, we cannot guarantee whether the results will be sustained in future code reviews.
}
    

\section{Conclusion}
\label{conclusion}

To understand the potential benefits of code reviews in identifying \securityIssue{s} from the \weakness{} aspect, we conducted an empirical case study to investigate the \securityConcern{s} raised during code reviews, their alignment with past \vulnerability{ies}, and their handling process. We manually validated and annotated the raised \securityConcern{s} into the software development perspective of the Common Weakness Enumeration (CWE-699). Then, we performed a qualitative analysis to investigate the alignment to the \knownV{} \vulnerability{ies} and the handling process of the raised concerns.

Based on the data from two large open-source projects, namely OpenSSL and PHP, our initial analysis indicates that \weakness{es} are 21 - 33.5 times more frequently discussed than \vulnerability{ies}. From our case study results, we found that code reviews can raise \securityConcern{s} from diverse \weakness{es}, accounting for 35 out of 40 categories in the CWE taxonomy. Some \securityConcern{s} are consistent across the two studied projects, such as authentication and privilege, API and behavior, and random number and cryptographic, while others are unique to the projects, such as direct security threats for OpenSSL and data validation for PHP.
The \weakness{es} in six weakness categories, e.g., memory errors, file handling, and numeric errors, are less frequently raised compared to the frequency of \knownV{} \vulnerability{ies}. Furthermore, there is a chance that \weakness{es} may slip through the code review process as only 37\% of \securityConcern{s} were fixed within the same code review process. The remaining cases resulted in no responses, dismissal, fixes in other areas, or unsuccessful resolutions. Our finding also highlighted an important shortcoming where 6\%-9\% of the code changes were merged although the \securityConcern{s} were not addressed.

Our study confirms that code reviews can identify \weakness{es} that may introduce \securityIssue{s}. Practitioners may focus on finding certain \weakness{es} when performing code reviews. However, checking all types of the \weakness{es} is not necessary.
Each project can prioritize the important \weakness{es} by analyzing their past \vulnerability{ies} and code review comments. Incomplete code reviews that contain \weakness{es} should be carefully monitored because they can slip through the code review process and may become \securityIssue{s} in the future. 
For future work, we encourage researchers to investigate the practical effectiveness of the \weakness-guided code review as well as develop a systematic approach to identify the project-specific \weakness{es} to assist the practitioners on a larger scale.


\section{Compliance with Ethical Standards}
\label{compliance}
The authors have no conflicts of interest to declare that are relevant to the content of this article.


\bibliographystyle{plainnat}
\bibliography{references}


\end{document}